\documentclass[floats,floatfix,showpacs,preprintnumbers,amssymb,prd,twocolumn,superscriptaddress,nofootinbib,nolongbibliography,reprint]{revtex4-1}

\usepackage{amssymb,amsmath,verbatim,mathtools,needspace,enumitem,etoolbox,graphicx,physics,microtype,afterpage,xspace,tabularx,lmodern,multirow,boldline, makecell, booktabs}
\usepackage{gensymb}
\usepackage{appendix}
\usepackage{tensor}
\usepackage{array}
\usepackage{siunitx}
\usepackage[normalem]{ulem}
\usepackage[dvipsnames, usenames]{xcolor}
\usepackage{xr-hyper}
\definecolor{linkcolor}{rgb}{0.0,0.3,0.5}
\usepackage[unicode, colorlinks=true, linkcolor=linkcolor, citecolor=linkcolor, filecolor=linkcolor, urlcolor=linkcolor, linktocpage, breaklinks]{hyperref}
\usepackage[all]{hypcap}
\usepackage[T1]{fontenc}
\usepackage[utf8]{inputenc}
\usepackage[usenames,dvipsnames]{xcolor}
\hypersetup{colorlinks=true,citecolor=magenta,linkcolor=violet,urlcolor=magenta}
\usepackage{multirow}

\definecolor{romared}{RGB}{142,0,28}

\newcommand{\be}{\begin{equation}}
\newcommand{\ee}{\end{equation}}

\def\be{\begin{equation}}
\def\ee{\end{equation}}
\newcommand{\beq}{\begin{eqnarray}}
\newcommand{\eeq}{\end{eqnarray}}

\usepackage{makecell}
\usepackage{soul}

\newcolumntype{Y}{>{\centering\arraybackslash}X}

\makeatletter
\newcommand*{\addFileDependency}[1]{
  \typeout{(#1)}
  \@addtofilelist{#1}
  \IfFileExists{#1}{}{\typeout{No file #1.}}
}
\makeatother

\newcommand*{\myexternaldocument}[1]{%
    \externaldocument{#1}%
    \addFileDependency{#1.tex}%
    \addFileDependency{#1.aux}%
}

\myexternaldocument{Supp}

\begin{document}
\title{Superradiant Instability of Magnetic Black Holes}

\author{David Pere\~niguez}
\email[]{david.pereniguez@nbi.ku.dk}
\affiliation{Niels Bohr International Academy, Niels Bohr Institute, Blegdamsvej 17, 2100 Copenhagen, Denmark}
\author{Marina de Amicis}
\affiliation{Niels Bohr International Academy, Niels Bohr Institute, Blegdamsvej 17, 2100 Copenhagen, Denmark}
\author{Richard Brito} 
\affiliation{CENTRA, Departamento de F\'{\i}sica, Instituto Superior T\'ecnico -- IST, Universidade de Lisboa -- UL, Avenida Rovisco Pais 1, 1049-001 Lisboa, Portugal}
\author{Rodrigo Panosso Macedo}
\affiliation{Niels Bohr International Academy, Niels Bohr Institute, Blegdamsvej 17, 2100 Copenhagen, Denmark}

\date{\today}

\begin{abstract}
Black hole superradiance has proven being very valuable in several realms of gravitational physics, and holds a promising discovery potential. In this paper, we consider the superradiant instability of magnetically-charged, rotating black holes and find a number of important differences with respect to neutral ones. Considering massive charged bosonic fields, we find that the instability timescale is much shorter, and this is true even if the black hole contains an order-one number of magnetic monopoles, or merely a single one, and possesses either low, moderate or large values of angular momentum. In particular, the instability is drastically faster than the radiative decay time of charged pions, potentially making it physically relevant. Furthermore, our analysis identifies the most unstable modes as a class of monopole spheroidal harmonics, that we dub north and south monopole modes, whose morphology is markedly different from the ones in standard superradiance since they extend along the rotational axis. For completeness, we also study the quasinormal mode spectrum and amplification factors of charged massless fields, finding no evidence of instabilities in that case.
\end{abstract}

\maketitle


\noindent{\bf{\em Introduction.}} Black holes stand as a remarkable prediction of theoretical physics. The observational validation of their existence either through precision astrometry \cite{2018A&A...618L..10G}, gravitational waves \cite{LIGOScientific:2016aoc}, large baseline interferometry \cite{EventHorizonTelescope:2019dse} or other astronomical observations~\cite{Narayan:2013gca}, mark a significant step forward in our ability to explore the hitherto hidden Universe \cite{Cardoso:2019rvt}. Beyond serving as optimal subjects for investigating the strong-field regime of gravitation, black holes also harbor immense potential for discovery. Over the past decade, the instabilities induced by black hole superradiance have proven instrumental in elucidating the structure of dark matter candidates \cite{Arvanitaki:2009fg,Brito:2015oca,Cardoso:2018tly}, exploring potential sources of gravitational waves \cite{Arvanitaki:2014wva,Brito:2017zvb,Aggarwal:2020olq} or electromagnetic signals~\cite{Rosa:2017ury,Ikeda:2018nhb,Ferraz:2020zgi,Caputo:2021efm,Siemonsen:2022ivj,Spieksma:2023vwl,Chakraborty:2024aug}, and even studying signatures of high-energy completions of General Relativity \cite{Alexander:2022avt,Richards:2023xsr}, among many other applications~\cite{Brito:2015oca}.

In this paper, we uncover a novel class of superradiant instability that may shed light on a long-standing problem in physics, that of the existence of magnetic monopoles in the Universe \cite{Preskill:1984gd}. Magnetic charges were considered back in 1904 by Thomson \cite{thomson_2009}, and recovered later on by Dirac \cite{Dirac:1931kp} who showed that the simultaneous existence of electric and magnetic charges $e$ and $P$ entails charge quantisation, as dictated by \textit{Dirac's quantisation condition},
\begin{equation}\label{DiracQuant}
    \frac{2Pe}{\hbar c}=\pm N\,, \ \ \ N=0,1,2,3...
\end{equation}
The fact that (primordial) magnetic monopoles are robust predictions of Grand Unified Theories \cite{tHooft:1974kcl,Polyakov:1974ek}, their ubiquity in String Theory \cite{Ortin:2015hya} and, of course, the fact that their existence would not contradict any known principle of Nature (quite the opposite!) has kept the research on magnetic monopoles an active field throughout the years, despite the lack of experimental evidence \cite{Giacomelli:2003yu,ATLAS:2023esy,Workman:2022ynf}. Such a lack of empirical corroboration could be due to the fact that magnetic monopoles are very heavy states, much more difficult to pair create than electric charges. In fact, monopoles can undergo gravitational collapse to form magnetic black holes \cite{Gibbons:1990um,Ortiz:1991eu,Lee:1991vy,Lee:1991qs}. An alternative formation channel of the latter is via monopole capture by primordial black holes, which provides an interesting alternative solution to the monopole problem in cosmology \cite{Stojkovic:2004hz}, or via monopole production in late stages of black hole evaporation \cite{Gibbons:1976sm,Profumo:2024fxq}. Since Schwinger pair-decay neutralisation is suppressed due to the large monopole's mass \cite{Gibbons:1975kk}, magnetic black holes retain their magnetic charge and do not undergo full evaporation by Hawking radiation, thus becoming long lived \cite{Maldacena:2020skw}. These reasons have triggered numerous works about magnetic black holes in the recent literature, concerning their implications for gravitational waves \cite{Liu:2020vsy,Liu:2020bag,Carullo:2021oxn,Chen:2022qvg,Pereniguez:2023wxf,DeFelice:2023rra}, dark matter and cosmology \cite{Bai:2019zcd,Kritos:2021nsf,Turner:1982ag,Kobayashi:2023ryr,Zhang:2023zmb,Wang:2023qxj}.

It is sensible to expect that primordial magnetic black holes contain an order-one number $N$ of monopoles and, consequently, that the influence of the latter in physical processes should be negligible. In this paper we show that even if the black hole contains a single monopole, massive bound states of charged bosonic fields in its vicinity suffer from a superradiant instability which surpasses significantly that of neutral Kerr black holes. Furthermore, this is true for either large, moderate or low (but non-zero) values of black hole angular momentum. This can be seen as a consequence of charge quantisation \eqref{DiracQuant}, which ``protects'' the electromagnetic interaction even if $P$ is arbitrarily small. In particular, for light primordial black holes with $M\lesssim 10^{12}$kg \cite{Carr:2020gox,Escriva:2022duf}, the instability timescale is drastically shorter than the mean lifetime of charged pions. This suggests that, unlike for neutral black holes \cite{Dolan:2007mj}, the instability could be physically realisable with standard model matter -- however, this also hinges on other physical aspects we discuss below. In addition, we show that the most unstable modes are a class of monopole harmonics \cite{Wu:1976ge}, that we dub north and south monopole modes, which extend along the rotational axis so their morphology differs largely from that in standard superradiance \cite{Konoplya:2013rxa}. However, the energy-momentum tensor of the cloud configuration is not axisymmetric, so it leads to electromagnetic and gravitational radiation (in a similar guise to other set ups, e.g. \cite{Arvanitaki:2010sy,Ikeda:2018nhb,Ferraz:2020zgi,East:2022ppo,Spieksma:2023vwl} and \cite{Arvanitaki:2014wva,Brito:2017zvb,Aggarwal:2020olq}). These results make magnetic black holes interesting sources in multi-messenger astronomy.

After reviewing briefly some aspects of magnetic monopoles, we describe our set up and present the results. We conclude by summarizing our findings and discussing future directions. In what follows we use natural units $G=c=\hbar=1$ and take Gauss's convention for the definition of charges.

\noindent{\bf{\em Charge quantisation and Monopole Harmonics.}} Physical systems composed by electric and magnetic charges exhibit a number of special features. Thomson \cite{thomson_2009} realised that a dipole constituted by an electric and magnetic charge pair, $e$ and $P$, known as a \textit{Thomson dipole}, carries angular momentum with magnitude given by the product of the charges. In his seminal work \cite{Dirac:1931kp}, Dirac established that the simultaneous existence of electric and magnetic charges entails charge quantisation, as dictated by \eqref{DiracQuant}. This, in turn, implies that the angular momentum of a Thomson dipole is quantised, in a way that is reminiscent of Bohr's postulate. Wu and Yang \cite{Wu:1976ge} showed that all these facts arise naturally when considering classical charged bosonic fields in the vicinity of a magnetic monopole, whose eigenstates are organised in the so-called monopole spherical harmonics. Heuristically, our aim is to replace the magnetic monopole by a rotating magnetic black hole. Being bosonic states, monopole spherical (in fact, spheroidal \cite{Semiz:1991kh}) modes are subject to superradiant amplification and, if endowed with a mass to bound them to the hole, they suffer from an instability, as we will show (the Penrose Process for charged particles in the vicinity of rotating magnetic black holes was considered in \cite{Dyson:2023ujk}). While sharing some resemblances with ordinary harmonics, monopole harmonics also possess some remarkable differences that is worth pointing out here (an extended discussion is given in Appendix \ref{A1}, for the interested reader). 

\begin{figure}[h!]
\includegraphics[width = 0.45\textwidth]{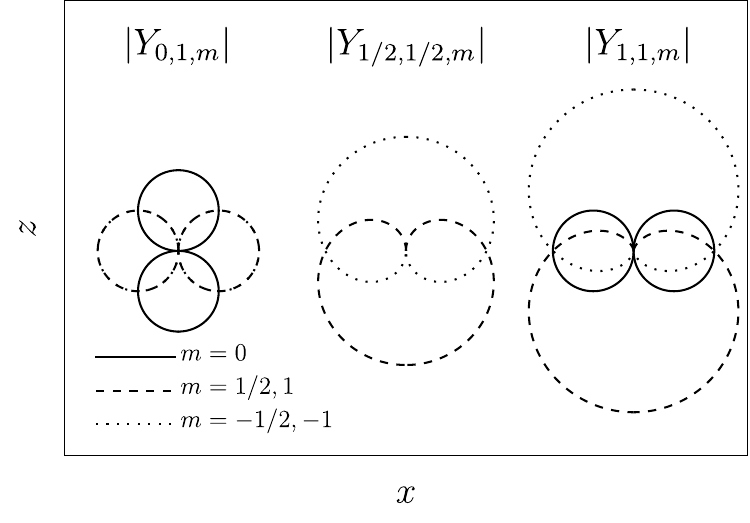}
\caption{Representation of monopole spherical harmonics, taking $r(\theta)=\lvert Y_{q,\ell,m}\rvert$. If $q\ne0$, north ($m=-q$) and south ($m=q$) monopole modes do not vanish at $\theta=0,\pi$, respectively, and will be the most superradiantly unstable modes. If $q=0$, the unstable modes will be $m=\pm1$, which vanish along the $z$ axis, as in the standard picture of superradiance. Note that the legend uses the same line types for north and south monopole modes in the different cases with $q\ne0$.}\label{MonoHarms}
\end{figure}

In three-dimensional Euclidean space, Dirac's monopole with charge $P$ is described by a $U(1)$ gauge field $\bold{A}$, and scalar fields $\varphi$ couple to it via the gauge-covariant derivative $\bold{D}=\boldsymbol{\nabla}+i e \bold{A}$, where $\boldsymbol{\nabla}$ is the Euclidean covariant derivative and $e$ is the charge of the field. Smoothness of this system as a principal $U(1)$-bundle entails Dirac's quantisation condition \eqref{DiracQuant}. Wu and Yang identified the following angular momentum operators
\begin{equation}\label{angmom}
        L_{l}=-i\epsilon_{ljk}x^{j}\bold{D}^{k}-q\frac{x^{l}}{r}\, ,\ \ \ \ (q\equiv-Pe)\, ,
\end{equation}
whose action on sections $\varphi$ is well defined (i.e.~it is covariant under gauge transformations). The first term in the definition \eqref{angmom} is the usual angular momentum operator for charged matter in the presence of a standard electromagnetic field. The presence of the second term, which is necessary in order to close the algebra of rotations, is an additional contribution due to the magnetic charge. Then, the \textit{monopole spherical harmonics} can be covariantly defined as smooth sections satisfying the eigenvalue problem
\begin{equation}\label{defY}
    L^{2}Y_{q,\ell,m}=\ell(\ell+1)Y_{q,\ell,m}\, , \ \ \ L_{z}Y_{q,\ell,m}=mY_{q,\ell,m}\, ,
\end{equation}
and standard representation theory gives \cite{Wu:1976ge}
\begin{align}
    \ell=\lvert q\rvert,\lvert q\rvert+1,...\, , \ \ \ \   m=-\ell, -\ell+1,...,  \ell\, .
\end{align}
A first surprising difference with respect to a magnetically neutral set up is that, for an odd number $N$ of magnetic monopoles, both $ \ell $ and $ m $ are half-integer and thus the field states form half-integer representations of rotations. A second, and most important difference, is the existence of modes that carry angular momentum in the $z$ direction, but do not vanish at the $z$ axis. Indeed, if $N\ne0$ it can be shown (see Appendix \ref{A1}) that modes with $m=-q$ and $m=q$ do not vanish at $\theta=0$ and $\theta=\pi$, respectively. We dub these modes the \textit{north and south monopole modes} (a few of them are illustrated in Fig.~\ref{MonoHarms}). In replacing the magnetic monopole by a rotating magnetic black hole, we will find that the north and south modes are the most amplified and unstable ones, thus leading to a novel class of black hole superradiance which is collimated along the rotational axis.\footnote{Strictly speaking, monopole spherical harmonics are smoothly deformed by the black hole rotation into monopole spheroidal harmonics, but the defining properties of north and south modes remain unchanged.} 
%


\noindent{\bf{\em Theory and set up.}} Scalar electrodynamics describes the interaction between electromagnetism and charged bosonic fields, such as pions, and its coupling to gravity is governed by the action
\begin{align} \notag
S[g,A,\varphi]&=\frac{1}{16\pi}\int d^{4}x\sqrt{-g}\left(R-F^{2}\right)\\ 
&-\frac{1}{2} \int d^{4}x\sqrt{-g}\left(D_{a}\bar{\varphi}D^{a}\varphi+\mu^{2}\bar{\varphi}\varphi\right)\, , \label{Action}
\end{align}
where $D_{a}=\nabla_{a}+ieA_{a}$ is the gauge-covariant derivative, $e$ and $\mu$ are the charge and the mass of the scalar field. Being a $U(1)$-gauge theory it has the local symmetry
\begin{equation}\label{eq:gauge}
\varphi \mapsto e^{i\alpha(x)}\varphi\, , \ \ \ \ \ A_{a}\mapsto A_{a}-\frac{1}{e}\ \nabla_{a}\alpha(x)\, ,
\end{equation}
generated by any real function $\alpha(x)$. From the no-hair conjectures (we are taking $\mu^{2}\geq0$ in this work) \cite{Bekenstein:1996pn}, one expects that quiescent, asymptotically flat black holes are accurately described by the dyonic Kerr--Newman spacetime, which has a vanishing scalar profile and is uniquely determined by its mass $M$, spin $J\equiv a M$, electric and magnetic charges $Q$ and $P$ (we review this solution in Appendix \ref{A2}). Due to the vanishing scalar background, first order scalar fluctuations decouple from the gravitational and electromagnetic ones. Therefore, to that order it suffices to consider solely the scalar field equation of motion. We seek for solutions with definite frequency $\omega$ and angular momentum $m$, which have the form
\begin{equation}
    \begin{aligned}\label{mode}
\varphi=e^{-i\omega t}e^{i m\phi}\Psi(r,\theta)=e^{-i\omega v}e^{i m\chi}\psi(r,\theta)\, ,
\end{aligned}
\end{equation}
expressed in Boyer--Lindquist $(t,r,\theta, \phi)$ and advanced Eddington--Finkelstein coordinates $(v,r,\theta, \chi)$, respectively. For the same reasons as in Dirac's monopole, regularity requires that the charge quantisation condition \eqref{DiracQuant} holds, and that $m\pm q$ take integer values.

The presence of an event horizon in the spacetime makes black holes dissipative systems. Consequently, the frequencies of free field oscillations have real and imaginary parts $\omega=\omega_{R}+i\omega_{I}$, the latter controlling the time-decay rate of the fluctuation. Using covariant phase-space methods \cite{Ortin:2022uxa, Dyson:2023ujk} it is possible to express the fluxes of energy, angular momentum and charge through the event horizon $H$ in the form
\begin{align}\label{DelE}
    \Delta E=&\left(\omega_{I}^{2}+\omega_{R}\omega_{*}\right)\int_{H}e^{2 \omega_{I} v}\lvert \psi \rvert^{2}\tilde{\boldsymbol{\epsilon}}\, ,\\ \label{DelJ}
    \Delta J=&m\omega_{*}\int_{H}e^{2 \omega_{I} v}\lvert \psi \rvert^{2}\tilde{\boldsymbol{\epsilon}}\,\, ,\\ \label{DelQ}
    \Delta Q=&-e \omega_{*}\int_{H}e^{2 \omega_{I} v}\lvert \psi \rvert^{2}\tilde{\boldsymbol{\epsilon}}\, ,
\end{align}
where $\tilde{\boldsymbol{\epsilon}}$ is the natural volume form on $H$ relative to the generator of the horizon. The superradiant threshold frequency $\omega_{*}$ is given by
\begin{equation}
    \omega_{*}=\omega_{R}-m\Omega+e\Phi\,,
\end{equation}
with $\Omega$ and $\Phi$ the black hole's angular velocity and electric potential, which are proportional to $a$ and $Q$, respectively (a detailed derivation of \eqref{DelE}-\eqref{DelQ} is given in the Appendix \ref{A3}). An important property of the flux formulas \eqref{DelE}-\eqref{DelQ} is that the integrals are positive-definite. Then, the signs of the pre factors serve as diagnostic quantities indicating whether the black hole is acreeting or releasing energy, angular momentum and charge. 

In the remainder of the paper we will restrict to purely magnetic black holes and set $Q=0$ (so $\Phi=0$, too). Then, solutions to the equations of motion depend on $a$, $M\mu$ and $e/\mu$ as in standard superradiance, and also on the number of magnetic monopoles contained in the hole, $N\equiv 2\lvert P e \rvert=2 \lvert q \rvert$. A solution is superradiant if it extracts energy from the black hole, $\Delta E<0$. From \eqref{DelE}-\eqref{DelQ} it follows that superradiant modes do also extract angular momentum, $\Delta J<0$, but can induce either positive or negative electric charge. In fact, for every solution \eqref{mode} there is another one, given by\footnote{This is not simply the complex conjugate of $\varphi$ because if $P\ne0$ then $\theta\mapsto \pi-\theta$ is an isometry of the metric but not of the field strength $F$, which picks up a sign $F\mapsto-F$.}
\begin{equation}\label{NtoS}
\tilde{\varphi}=e^{i\bar{\omega} v}e^{-i m\chi}\bar{\psi}(r,\pi-\theta)\, ,
\end{equation}
which induces the same amount of electric charge but with opposite sign, and that is its ``mirror image'' with respect to the equatorial plane $\theta=\pi/2$. North and south monopole modes correspond under this map (notice that, in particular, the map sends $(\omega_{R},\omega_{I})\mapsto(-\omega_{R},\omega_{I})$ and $m\mapsto-m$). Superradiant instabilities arise as massive bound states in the superradiant regime that grow exponentially in time~\cite{Brito:2015oca}. More precisely, the general asymptotic behaviour of the solutions is
\begin{equation}\label{infty}
    \varphi\sim e^{\pm  \sqrt{\mu^{2}-\omega^{2}}r}\, , \ \ \ \ r\to\infty\, ,
\end{equation}
and bound states are defined as exponentially-supressed massive modes, so they only exhibit the decaying behaviour of \eqref{infty}. This asymptotic boundary condition, together with regularity of the fluctuation on and outside the event horizon\footnote{One should impose regularity of $\varphi$ as a $U(1)$ section, that is, the field must be regular when expressed in the coordinates and the gauges where $g_{ab}$ and $A_{a}$ are regular. This is reviewed in detail in Appendix \ref{A4}.} defines a characteristic value problem for the frequencies. The confining effect due to the field's mass, combined with superradiant amplification leads to solutions with $\omega_{I}>0$. The rest of the paper is devoted to the study of these superradiant instabilities. Given that the equation for the field separates, $\Psi(r,\theta)=R(r)S(\theta)$, the frequencies of bound states can be found using a standard Leaver method \cite{Leaver:1985ax,Cardoso:2005vk,Dolan:2007mj} that we review in the Appendix \ref{A4}.


\noindent{\bf{\em Results.}} To illustrate the features of the instability, we shall consider the following range of parameters, and comment on this choice later on. We fix the charge-to-mass ratio of the field to be that of charged pions $\pi^{\pm}$, $e/\mu=7.4\times 10^{18}$ in natural units \cite{Workman:2022ynf}, and restrict to the regime $M\mu\lesssim 1$ which for pions corresponds to a mass range relevant for light primordial black holes $M\lesssim 10^{12}$kg \cite{Carr:2020gox,Escriva:2022duf}. We allow the black hole to carry up to three magnetic monopoles, $N=0,1,2,3$, and restrict ourselves to the most unstable modes. As we will show, these are the north and south monopole modes $m=\pm q=\pm N/2$ with $\ell=\lvert q\rvert$ for $N\ne0$, and of course the $\ell=m=1$ mode  for $N=0$ (the case of a neutral Kerr black hole). The frequencies displayed for moderate and large values of black hole spin in Fig.~\ref{omegas} show that the dependence in $M\mu$ is qualitatively similar to that of neutral black holes. However, it is quite surprising that the addition of just a few magnetic monopoles, which corresponds to a negligible charge-to-mass ratio $P/M\sim 10^{-19}$, induces finite changes in the frequency spectrum. This fact is even more severe in the unstable regime, highlighted in Fig.~\ref{instabs} after zooming into the left panel of Fig.~\ref{omegas}. Strikingly, we find that the instability timescale $\tau=1/\omega_{I}$ can be decreased by as much as two orders of magnitude (e.g.~comparing the $N=0$ and $N=2$ cases).
\begin{widetext}
\begin{center}
    \begin{figure}[h!]
       \includegraphics[width = 0.95\textwidth]{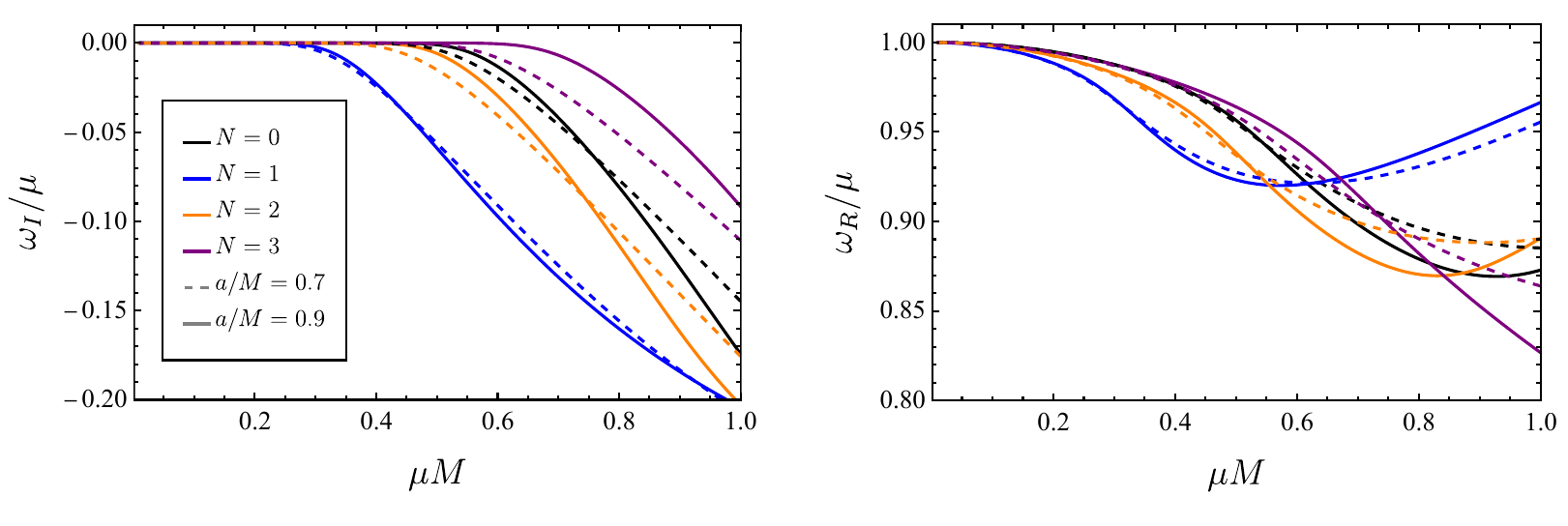}
\caption{Real and imaginary parts of the frequency of the bound state modes of a charged scalar field, with the charge-to-mass ratio of charged pions $\pi^{\pm}$, on a magnetic black hole with $N=0,1,2,3$ monopoles spinning at $a/M=0.7$ and $a/M=0.9$. For $N=0$ (which corresponds to the neutral Kerr black hole) we show the mode $\ell=m=1$, while for $N=1,2,3$ we show the north monopole modes $\ell=m=-q=N/2$. The frequencies of the south monopole modes are obtained from the north ones as $(\omega_{R},\omega_{I})\mapsto(-\omega_{R},\omega_{I})$.}\label{omegas}
\end{figure}
\end{center}
\end{widetext}
To discuss the physical viability of our instability in this example, consider the black hole in isolation, free of matter environments. In that case, the superradiant instability can only be triggered by spontaneous emission of the scalar field \cite{Gibbons:1975kk,Page:1976df,Page:1976ki,Page:1977um}. To estimate whether this can happen, we consider the electric field sourced by the rotating magnetic black hole at the axes, $\theta=0,\pi$, where the north and south modes live.\footnote{Explicitly, the electric field is $E_{a}=k^{b}F_{ab}$, where $k^{a}=t^{a}+\Omega \phi^{a}$ is the horizon's generator \cite{Carter:1973rla}.} It is given by $e E_{r}\sim 0.15 (a/M)N/M^{2}$, and for $N\sim1$, natural values of $a/M$, and in the mass range $M\mu\lesssim1$ one has that $e E_{r}\gtrsim \mu^{2}$. This is the condition under which Schwinger’s pair production becomes efficient, and it holds precisely in the regime of the parameters where we find our instability. While this suggests that the instability could be triggered by spontaneous emission, in order to evolve enough in time its timescale needs to be shorter than the radiative-decay lifetime of the particles -- charged pions in our case. As an example, consider a black hole with spin $a/M=0.9$ that contains $N=3$ magnetic monopoles. From Fig.~\ref{instabs} we find an instability that is more severe at $M\mu\sim0.45$. For charged pions $\mu_{\pi^{\pm}}\sim140$ MeV/c$^2$ \cite{Workman:2022ynf}, which corresponds to a black hole mass of $M\sim 8\times10^{11}$kg, and the instability timescale is $\tau\sim 1.5\times 10^{-18} $s. This is ten orders of magnitude shorter than the mean lifetime of charged pions, $\tau_{\pi^{\pm}}\sim 2.6\times 10^{-8}$s \cite{Workman:2022ynf}, so the instability would safely evolve in time without being quenched by radiative particle decay. It is also worth noticing that the instability is a hundred times faster than that for a neutral black hole with the same spin ($\tau\sim 10^{-16}$s), even though these only differ by three monopoles. In fact, the separation between timescales is large enough so that the instability would set in even for low values of the spin. For a black hole containing $N=1$ monopole and spinning at $a/M=0.3$, we find that the instability peaks at a black hole mass $M\sim 6\times 10^{10}$kg and its timescale is $\tau\sim 1.5\times 10^{-13}$s, still much shorter than $\tau_{\pi^{\pm}}$.

We should emphasize, though, that these are only first estimates and establishing robustly that the instability would realise in nature requires addressing more aspects. On the one hand, to be rigorous Schwinger's formula is accurate in describing particle creation near black holes only in the regime $M\gg\mu$. Thus, the efficiency of spontaneous emission as a trigger for the instability should be confirmed more accurately. Besides, the electric field would also emit other lighter particles, like electrons, which may lead to screening effects. Being fermions, Pauli's exclusion principle prevents them from accumulating via a superradiant instability, as is well-known \cite{Brito:2015oca}, but they may still yield important effects. An analysis of these issues requires computing the spectrum of excitations by magnetic black holes in the entire standard model, a fact that has been addressed only partially in restricted scenarios \cite{Gibbons:1976sm}.

\begin{figure}[t!]
\includegraphics[width = 0.3\textwidth]{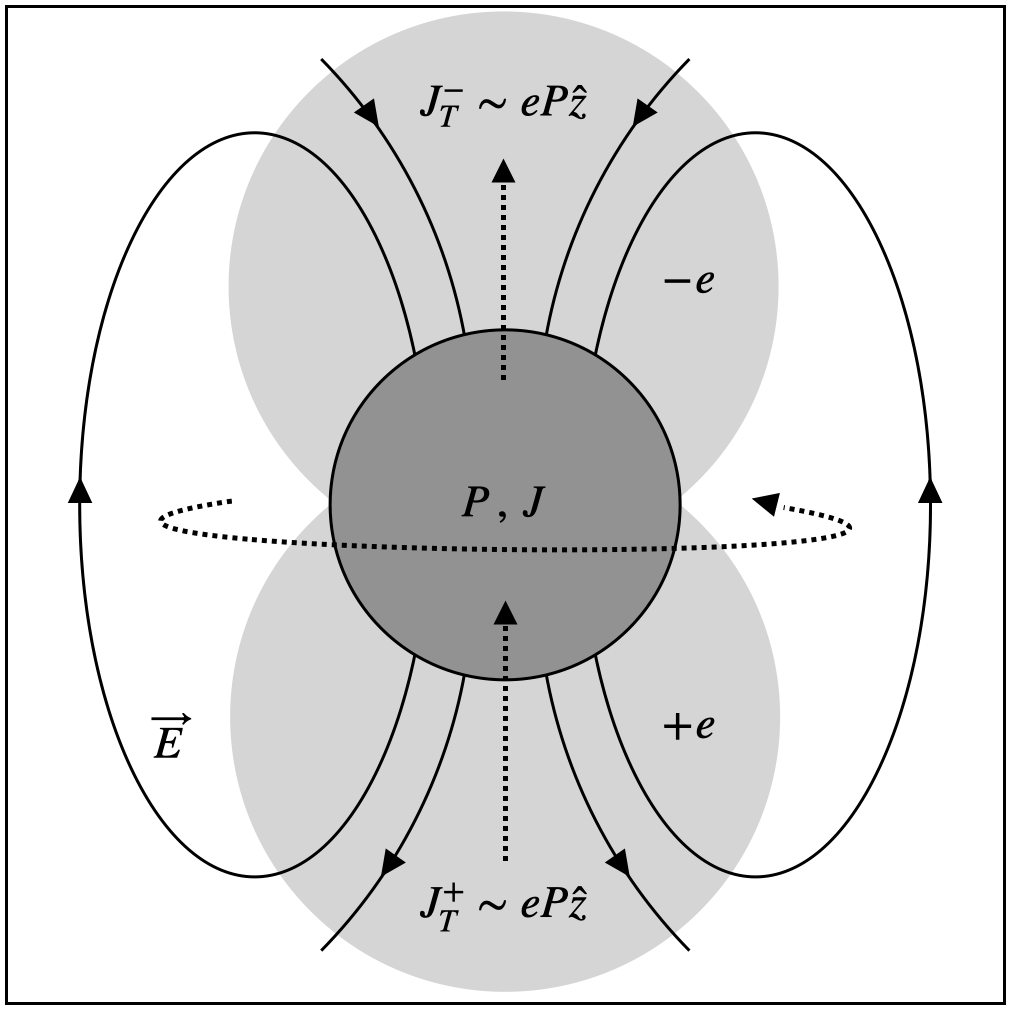}
\caption{Sketch of the superradiant instability of electrically charged matter in the presence of a rotating magnetic black hole, taking $J,P$, and $e$ positive.}\label{Pic}
\end{figure}

We conclude this section by noting that north and south monopole modes $m=\pm q$ are the most unstable ones. These exhibit a morphology that is markedly different from the usual superradiantly unstable modes, since they extend along the north and south components of the rotational axis (see Fig.~\ref{MonoHarms} and Fig.~\ref{Pic}). We find that higher modes $\ell>\lvert q \rvert$ have longer instability timescales. For completeness, in Appendix \ref{A5} we elaborate on other quantities that help characterising the instability, such as the fluxes of energy, angular momentum and charge through the horizon, as well as the superradiant amplification factors. We have also studied the quasinormal mode spectrum of massless charged fields in the vicinity of a rotating, magnetic black hole, and reported the results in Appendix \ref{A6}. In contrast with massive bound states, we find no trace of instabilities. 
\begin{figure}[t!]
\includegraphics[width = 0.48\textwidth]{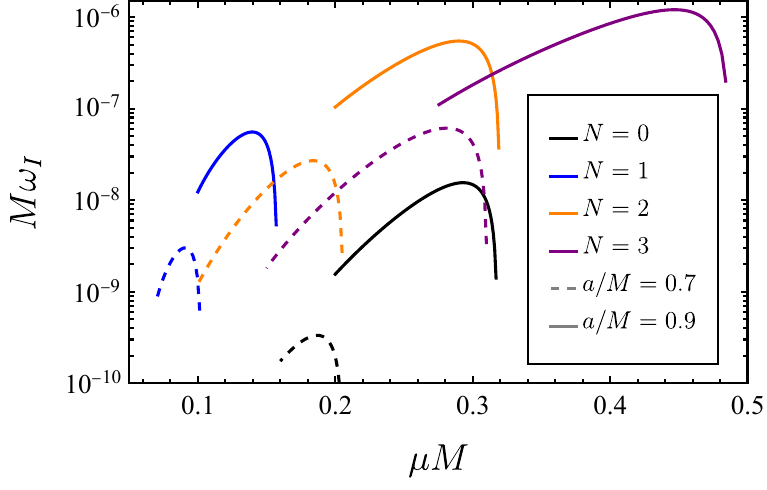}
\caption{Superradiant instability of the bound state modes in Fig.~\ref{omegas}. The same results hold also for the south monopole modes.}\label{instabs}
\end{figure}
%

\noindent{\bf{\em Discussion.}} Our results establish the following picture for the superradiant instability of magnetic black holes. As the hole rotates, it sources a dipolar electric field and electrically charged matter in its vicinity is superradiantly amplified. If possessing a mass that binds it to the hole, such matter is unstable to forming two spheroidal clouds. These clouds grow along the north and south components of the rotational axis, have opposite charge signs, and their morphology is that of north and south monopole spheroidal modes. The system formed by the black hole and each of the clouds can be seen as a Thomson dipole, whose angular momentum is $J_{T}= - q_{\text{cloud}} P  \hat{r}$, where $q_{\text{cloud}}$ is the charge of the cloud and $\hat{r}$ points from the black hole towards the cloud \cite{thomson_2009,Dyson:2023ujk}. While the hole rotates, energy and angular momentum are being transferred into the formation of both effective Thomson dipoles. Each cloud induces the same flux of electric charge into the hole, but with opposite sign, so the net amount of electric black hole charge remains zero. The entire process is summarised in Fig.~\ref{Pic}. 

Analogues of this instability are expected to occur in magnetic black holes in several other contexts, ranging from non-Abelian gauge theories (such as rotating versions of \cite{Ortiz:1991eu,Lee:1991qs,Lee:1991vy}) to some supergravity theories \cite{Lozano-Tellechea:1999lwm} and other models \cite{DeFelice:2024eoj}, and we expect its main characteristics to remain. Namely, that variations in the number of monopoles modify significantly the instability spectrum even if the hole's magnetic charge-to-mass ratio remains negligible, and that the unstable modes grow along the rotational axis, via a transfer of angular momentum, resembling a pair of Thomson dipoles.
In addition, we have demonstrated with an example that the instability also yields interesting phenomenology. Employing scalar electrodynamics to describe charged pions, we have shown that rotating black holes with $M\lesssim 10^{12}$kg hosting an order-one number of magnetic monopoles $N\sim1$ are unstable to forming charged pion clouds. This is, in fact, quite reminiscent of the pion condensates formed when magnetic fields and rotation coexist \cite{Liu:2017spl}. A main phenomenological caveat, however, are the black hole masses considered. Primordial black holes with initial mass $M\lesssim 10^{12}$kg may have evaporated by now \cite{Page:1976df,Page:1976ki,Page:1977um}, so their existence hinges on less conventional mechanisms (such as primordial black hole mergers \cite{Rice:2017avg,Franciolini:2022htd}). Studying in detail formation channels of these black holes is left to future research.

Tracking the evolution of the instability numerically would reveal the precise physics of the transient, which may involve metastable hairy solutions \cite{Herdeiro:2014goa}. Besides emitting gravitational and electromagnetic waves \cite{Aggarwal:2020olq}, it is possible that the instability results in a violent annihilation of the oppositely-charged clouds. If so, it would be interesting to regard this phenomenon as a potential source of highly-energetic cosmic rays coming from void regions \cite{Bhattacharjee:1999mup,TelescopeArray:2023sbd}. Finally, departing form the parameters concerning matter fields in the Standard Model, e.g. considering ultralight bosonic millicharged fields, the instability range can cover super-massive black holes. In particular, it is a curious coincidence that the morphology of the unstable north and south monopole modes is quite similar to the Fermi bubbles of our Galaxy, and their much larger X-ray analogue \cite{Predehl:2020kyq}. An exploration of these research directions requires an accurate analysis of the evolution of the instability.


\noindent{\bf{\em Acknowledgements.}} We thank Diego Blas, Vitor Cardoso, Roberto Emparan, Antonio de Felice, Jaume Garriga, David Mateos, Tomás Ortín, Paolo Pani and Shinji Tsujikawa for insightful conversations. We acknowledge financial support by the VILLUM Foundation (grant no. VIL37766) and the DNRF Chair program (grant no. DNRF162) by the Danish National Research Foundation. This project has received funding from the European Union’s Horizon 2020 research and innovation programme under the Marie Sklodowska-Curie
grant agreement No 101007855 and No 101007855. R.B. acknowledges financial support provided by FCT – Fundação para a Ciência e a Tecnologia, I.P., under the Scientific Employment Stimulus -- Individual Call -- Grant No. 2020.00470.CEECIND and under Project No. 2022.01324.PTDC. 

\bibliography{refMAIN}

\begin{thebibliography}{86}%
\makeatletter
\providecommand \@ifxundefined [1]{%
 \@ifx{#1\undefined}
}%
\providecommand \@ifnum [1]{%
 \ifnum #1\expandafter \@firstoftwo
 \else \expandafter \@secondoftwo
 \fi
}%
\providecommand \@ifx [1]{%
 \ifx #1\expandafter \@firstoftwo
 \else \expandafter \@secondoftwo
 \fi
}%
\providecommand \natexlab [1]{#1}%
\providecommand \enquote  [1]{``#1''}%
\providecommand \bibnamefont  [1]{#1}%
\providecommand \bibfnamefont [1]{#1}%
\providecommand \citenamefont [1]{#1}%
\providecommand \href@noop [0]{\@secondoftwo}%
\providecommand \href [0]{\begingroup \@sanitize@url \@href}%
\providecommand \@href[1]{\@@startlink{#1}\@@href}%
\providecommand \@@href[1]{\endgroup#1\@@endlink}%
\providecommand \@sanitize@url [0]{\catcode `\\12\catcode `\$12\catcode
  `\&12\catcode `\#12\catcode `\^12\catcode `\_12\catcode `\%12\relax}%
\providecommand \@@startlink[1]{}%
\providecommand \@@endlink[0]{}%
\providecommand \url  [0]{\begingroup\@sanitize@url \@url }%
\providecommand \@url [1]{\endgroup\@href {#1}{\urlprefix }}%
\providecommand \urlprefix  [0]{URL }%
\providecommand \Eprint [0]{\href }%
\providecommand \doibase [0]{http://dx.doi.org/}%
\providecommand \selectlanguage [0]{\@gobble}%
\providecommand \bibinfo  [0]{\@secondoftwo}%
\providecommand \bibfield  [0]{\@secondoftwo}%
\providecommand \translation [1]{[#1]}%
\providecommand \BibitemOpen [0]{}%
\providecommand \bibitemStop [0]{}%
\providecommand \bibitemNoStop [0]{.\EOS\space}%
\providecommand \EOS [0]{\spacefactor3000\relax}%
\providecommand \BibitemShut  [1]{\csname bibitem#1\endcsname}%
\let\auto@bib@innerbib\@empty
\bibitem [{\citenamefont {Abuter}\ \emph {et~al.}(2018)\citenamefont {Abuter}
  \emph {et~al.}}]{2018A&A...618L..10G}%
  \BibitemOpen
  \bibfield  {author} {\bibinfo {author} {\bibfnamefont {R.}~\bibnamefont
  {Abuter}} \emph {et~al.} (\bibinfo {collaboration} {GRAVITY Collaboration}),\
  }\href {\doibase 10.1051/0004-6361/201834294} {\bibfield  {journal} {\bibinfo
   {journal} {Astronomy and Astrophysics}\ }\textbf {\bibinfo {volume} {618}},\
  \bibinfo {eid} {L10} (\bibinfo {year} {2018})},\ \Eprint
  {http://arxiv.org/abs/1810.12641} {arXiv:1810.12641 [astro-ph.GA]}
  \BibitemShut {NoStop}%
\bibitem [{\citenamefont {Abbott}\ \emph {et~al.}(2016)\citenamefont {Abbott}
  \emph {et~al.}}]{LIGOScientific:2016aoc}%
  \BibitemOpen
  \bibfield  {author} {\bibinfo {author} {\bibfnamefont {B.~P.}\ \bibnamefont
  {Abbott}} \emph {et~al.} (\bibinfo {collaboration} {LIGO Scientific,
  Virgo}),\ }\href {\doibase 10.1103/PhysRevLett.116.061102} {\bibfield
  {journal} {\bibinfo  {journal} {Phys. Rev. Lett.}\ }\textbf {\bibinfo
  {volume} {116}},\ \bibinfo {pages} {061102} (\bibinfo {year} {2016})},\
  \Eprint {http://arxiv.org/abs/1602.03837} {arXiv:1602.03837 [gr-qc]}
  \BibitemShut {NoStop}%
\bibitem [{\citenamefont {Akiyama}\ \emph {et~al.}(2019)\citenamefont {Akiyama}
  \emph {et~al.}}]{EventHorizonTelescope:2019dse}%
  \BibitemOpen
  \bibfield  {author} {\bibinfo {author} {\bibfnamefont {K.}~\bibnamefont
  {Akiyama}} \emph {et~al.} (\bibinfo {collaboration} {Event Horizon
  Telescope}),\ }\href {\doibase 10.3847/2041-8213/ab0ec7} {\bibfield
  {journal} {\bibinfo  {journal} {Astrophys. J. Lett.}\ }\textbf {\bibinfo
  {volume} {875}},\ \bibinfo {pages} {L1} (\bibinfo {year} {2019})},\ \Eprint
  {http://arxiv.org/abs/1906.11238} {arXiv:1906.11238 [astro-ph.GA]}
  \BibitemShut {NoStop}%
\bibitem [{\citenamefont {Narayan}\ and\ \citenamefont
  {McClintock}(2013)}]{Narayan:2013gca}%
  \BibitemOpen
  \bibfield  {author} {\bibinfo {author} {\bibfnamefont {R.}~\bibnamefont
  {Narayan}}\ and\ \bibinfo {author} {\bibfnamefont {J.~E.}\ \bibnamefont
  {McClintock}},\ }\href@noop {} {\  (\bibinfo {year} {2013})},\ \Eprint
  {http://arxiv.org/abs/1312.6698} {arXiv:1312.6698 [astro-ph.HE]} \BibitemShut
  {NoStop}%
\bibitem [{\citenamefont {Cardoso}\ and\ \citenamefont
  {Pani}(2019)}]{Cardoso:2019rvt}%
  \BibitemOpen
  \bibfield  {author} {\bibinfo {author} {\bibfnamefont {V.}~\bibnamefont
  {Cardoso}}\ and\ \bibinfo {author} {\bibfnamefont {P.}~\bibnamefont {Pani}},\
  }\href {\doibase 10.1007/s41114-019-0020-4} {\bibfield  {journal} {\bibinfo
  {journal} {Living Rev. Rel.}\ }\textbf {\bibinfo {volume} {22}},\ \bibinfo
  {pages} {4} (\bibinfo {year} {2019})},\ \Eprint
  {http://arxiv.org/abs/1904.05363} {arXiv:1904.05363 [gr-qc]} \BibitemShut
  {NoStop}%
\bibitem [{\citenamefont {Arvanitaki}\ \emph {et~al.}(2010)\citenamefont
  {Arvanitaki}, \citenamefont {Dimopoulos}, \citenamefont {Dubovsky},
  \citenamefont {Kaloper},\ and\ \citenamefont
  {March-Russell}}]{Arvanitaki:2009fg}%
  \BibitemOpen
  \bibfield  {author} {\bibinfo {author} {\bibfnamefont {A.}~\bibnamefont
  {Arvanitaki}}, \bibinfo {author} {\bibfnamefont {S.}~\bibnamefont
  {Dimopoulos}}, \bibinfo {author} {\bibfnamefont {S.}~\bibnamefont
  {Dubovsky}}, \bibinfo {author} {\bibfnamefont {N.}~\bibnamefont {Kaloper}}, \
  and\ \bibinfo {author} {\bibfnamefont {J.}~\bibnamefont {March-Russell}},\
  }\href {\doibase 10.1103/PhysRevD.81.123530} {\bibfield  {journal} {\bibinfo
  {journal} {Phys. Rev. D}\ }\textbf {\bibinfo {volume} {81}},\ \bibinfo
  {pages} {123530} (\bibinfo {year} {2010})},\ \Eprint
  {http://arxiv.org/abs/0905.4720} {arXiv:0905.4720 [hep-th]} \BibitemShut
  {NoStop}%
\bibitem [{\citenamefont {Brito}\ \emph {et~al.}(2015)\citenamefont {Brito},
  \citenamefont {Cardoso},\ and\ \citenamefont {Pani}}]{Brito:2015oca}%
  \BibitemOpen
  \bibfield  {author} {\bibinfo {author} {\bibfnamefont {R.}~\bibnamefont
  {Brito}}, \bibinfo {author} {\bibfnamefont {V.}~\bibnamefont {Cardoso}}, \
  and\ \bibinfo {author} {\bibfnamefont {P.}~\bibnamefont {Pani}},\ }\href
  {\doibase 10.1007/978-3-319-19000-6} {\bibfield  {journal} {\bibinfo
  {journal} {Lect. Notes Phys.}\ }\textbf {\bibinfo {volume} {906}},\ \bibinfo
  {pages} {pp.1} (\bibinfo {year} {2015})},\ \Eprint
  {http://arxiv.org/abs/1501.06570} {arXiv:1501.06570 [gr-qc]} \BibitemShut
  {NoStop}%
\bibitem [{\citenamefont {Cardoso}\ \emph {et~al.}(2018)\citenamefont
  {Cardoso}, \citenamefont {Dias}, \citenamefont {Hartnett}, \citenamefont
  {Middleton}, \citenamefont {Pani},\ and\ \citenamefont
  {Santos}}]{Cardoso:2018tly}%
  \BibitemOpen
  \bibfield  {author} {\bibinfo {author} {\bibfnamefont {V.}~\bibnamefont
  {Cardoso}}, \bibinfo {author} {\bibfnamefont {O.~J.~C.}\ \bibnamefont
  {Dias}}, \bibinfo {author} {\bibfnamefont {G.~S.}\ \bibnamefont {Hartnett}},
  \bibinfo {author} {\bibfnamefont {M.}~\bibnamefont {Middleton}}, \bibinfo
  {author} {\bibfnamefont {P.}~\bibnamefont {Pani}}, \ and\ \bibinfo {author}
  {\bibfnamefont {J.~E.}\ \bibnamefont {Santos}},\ }\href {\doibase
  10.1088/1475-7516/2018/03/043} {\bibfield  {journal} {\bibinfo  {journal}
  {JCAP}\ }\textbf {\bibinfo {volume} {03}},\ \bibinfo {pages} {043} (\bibinfo
  {year} {2018})},\ \Eprint {http://arxiv.org/abs/1801.01420} {arXiv:1801.01420
  [gr-qc]} \BibitemShut {NoStop}%
\bibitem [{\citenamefont {Arvanitaki}\ \emph {et~al.}(2015)\citenamefont
  {Arvanitaki}, \citenamefont {Baryakhtar},\ and\ \citenamefont
  {Huang}}]{Arvanitaki:2014wva}%
  \BibitemOpen
  \bibfield  {author} {\bibinfo {author} {\bibfnamefont {A.}~\bibnamefont
  {Arvanitaki}}, \bibinfo {author} {\bibfnamefont {M.}~\bibnamefont
  {Baryakhtar}}, \ and\ \bibinfo {author} {\bibfnamefont {X.}~\bibnamefont
  {Huang}},\ }\href {\doibase 10.1103/PhysRevD.91.084011} {\bibfield  {journal}
  {\bibinfo  {journal} {Phys. Rev. D}\ }\textbf {\bibinfo {volume} {91}},\
  \bibinfo {pages} {084011} (\bibinfo {year} {2015})},\ \Eprint
  {http://arxiv.org/abs/1411.2263} {arXiv:1411.2263 [hep-ph]} \BibitemShut
  {NoStop}%
\bibitem [{\citenamefont {Brito}\ \emph {et~al.}(2017)\citenamefont {Brito},
  \citenamefont {Ghosh}, \citenamefont {Barausse}, \citenamefont {Berti},
  \citenamefont {Cardoso}, \citenamefont {Dvorkin}, \citenamefont {Klein},\
  and\ \citenamefont {Pani}}]{Brito:2017zvb}%
  \BibitemOpen
  \bibfield  {author} {\bibinfo {author} {\bibfnamefont {R.}~\bibnamefont
  {Brito}}, \bibinfo {author} {\bibfnamefont {S.}~\bibnamefont {Ghosh}},
  \bibinfo {author} {\bibfnamefont {E.}~\bibnamefont {Barausse}}, \bibinfo
  {author} {\bibfnamefont {E.}~\bibnamefont {Berti}}, \bibinfo {author}
  {\bibfnamefont {V.}~\bibnamefont {Cardoso}}, \bibinfo {author} {\bibfnamefont
  {I.}~\bibnamefont {Dvorkin}}, \bibinfo {author} {\bibfnamefont
  {A.}~\bibnamefont {Klein}}, \ and\ \bibinfo {author} {\bibfnamefont
  {P.}~\bibnamefont {Pani}},\ }\href {\doibase 10.1103/PhysRevD.96.064050}
  {\bibfield  {journal} {\bibinfo  {journal} {Phys. Rev. D}\ }\textbf {\bibinfo
  {volume} {96}},\ \bibinfo {pages} {064050} (\bibinfo {year} {2017})},\
  \Eprint {http://arxiv.org/abs/1706.06311} {arXiv:1706.06311 [gr-qc]}
  \BibitemShut {NoStop}%
\bibitem [{\citenamefont {Aggarwal}\ \emph {et~al.}(2021)\citenamefont
  {Aggarwal} \emph {et~al.}}]{Aggarwal:2020olq}%
  \BibitemOpen
  \bibfield  {author} {\bibinfo {author} {\bibfnamefont {N.}~\bibnamefont
  {Aggarwal}} \emph {et~al.},\ }\href {\doibase 10.1007/s41114-021-00032-5}
  {\bibfield  {journal} {\bibinfo  {journal} {Living Rev. Rel.}\ }\textbf
  {\bibinfo {volume} {24}},\ \bibinfo {pages} {4} (\bibinfo {year} {2021})},\
  \Eprint {http://arxiv.org/abs/2011.12414} {arXiv:2011.12414 [gr-qc]}
  \BibitemShut {NoStop}%
\bibitem [{\citenamefont {Rosa}\ and\ \citenamefont
  {Kephart}(2018)}]{Rosa:2017ury}%
  \BibitemOpen
  \bibfield  {author} {\bibinfo {author} {\bibfnamefont {J.~a.~G.}\
  \bibnamefont {Rosa}}\ and\ \bibinfo {author} {\bibfnamefont {T.~W.}\
  \bibnamefont {Kephart}},\ }\href {\doibase 10.1103/PhysRevLett.120.231102}
  {\bibfield  {journal} {\bibinfo  {journal} {Phys. Rev. Lett.}\ }\textbf
  {\bibinfo {volume} {120}},\ \bibinfo {pages} {231102} (\bibinfo {year}
  {2018})},\ \Eprint {http://arxiv.org/abs/1709.06581} {arXiv:1709.06581
  [gr-qc]} \BibitemShut {NoStop}%
\bibitem [{\citenamefont {Ikeda}\ \emph {et~al.}(2019)\citenamefont {Ikeda},
  \citenamefont {Brito},\ and\ \citenamefont {Cardoso}}]{Ikeda:2018nhb}%
  \BibitemOpen
  \bibfield  {author} {\bibinfo {author} {\bibfnamefont {T.}~\bibnamefont
  {Ikeda}}, \bibinfo {author} {\bibfnamefont {R.}~\bibnamefont {Brito}}, \ and\
  \bibinfo {author} {\bibfnamefont {V.}~\bibnamefont {Cardoso}},\ }\href
  {\doibase 10.1103/PhysRevLett.122.081101} {\bibfield  {journal} {\bibinfo
  {journal} {Phys. Rev. Lett.}\ }\textbf {\bibinfo {volume} {122}},\ \bibinfo
  {pages} {081101} (\bibinfo {year} {2019})},\ \Eprint
  {http://arxiv.org/abs/1811.04950} {arXiv:1811.04950 [gr-qc]} \BibitemShut
  {NoStop}%
\bibitem [{\citenamefont {Ferraz}\ \emph {et~al.}(2022)\citenamefont {Ferraz},
  \citenamefont {Kephart},\ and\ \citenamefont {Rosa}}]{Ferraz:2020zgi}%
  \BibitemOpen
  \bibfield  {author} {\bibinfo {author} {\bibfnamefont {P.~B.}\ \bibnamefont
  {Ferraz}}, \bibinfo {author} {\bibfnamefont {T.~W.}\ \bibnamefont {Kephart}},
  \ and\ \bibinfo {author} {\bibfnamefont {J.~a.~G.}\ \bibnamefont {Rosa}},\
  }\href {\doibase 10.1088/1475-7516/2022/07/026} {\bibfield  {journal}
  {\bibinfo  {journal} {JCAP}\ }\textbf {\bibinfo {volume} {07}},\ \bibinfo
  {pages} {026} (\bibinfo {year} {2022})},\ \Eprint
  {http://arxiv.org/abs/2004.11303} {arXiv:2004.11303 [gr-qc]} \BibitemShut
  {NoStop}%
\bibitem [{\citenamefont {Caputo}\ \emph {et~al.}(2021)\citenamefont {Caputo},
  \citenamefont {Witte}, \citenamefont {Blas},\ and\ \citenamefont
  {Pani}}]{Caputo:2021efm}%
  \BibitemOpen
  \bibfield  {author} {\bibinfo {author} {\bibfnamefont {A.}~\bibnamefont
  {Caputo}}, \bibinfo {author} {\bibfnamefont {S.~J.}\ \bibnamefont {Witte}},
  \bibinfo {author} {\bibfnamefont {D.}~\bibnamefont {Blas}}, \ and\ \bibinfo
  {author} {\bibfnamefont {P.}~\bibnamefont {Pani}},\ }\href {\doibase
  10.1103/PhysRevD.104.043006} {\bibfield  {journal} {\bibinfo  {journal}
  {Phys. Rev. D}\ }\textbf {\bibinfo {volume} {104}},\ \bibinfo {pages}
  {043006} (\bibinfo {year} {2021})},\ \Eprint
  {http://arxiv.org/abs/2102.11280} {arXiv:2102.11280 [hep-ph]} \BibitemShut
  {NoStop}%
\bibitem [{\citenamefont {Siemonsen}\ \emph {et~al.}(2023)\citenamefont
  {Siemonsen}, \citenamefont {Mondino}, \citenamefont {Egana-Ugrinovic},
  \citenamefont {Huang}, \citenamefont {Baryakhtar},\ and\ \citenamefont
  {East}}]{Siemonsen:2022ivj}%
  \BibitemOpen
  \bibfield  {author} {\bibinfo {author} {\bibfnamefont {N.}~\bibnamefont
  {Siemonsen}}, \bibinfo {author} {\bibfnamefont {C.}~\bibnamefont {Mondino}},
  \bibinfo {author} {\bibfnamefont {D.}~\bibnamefont {Egana-Ugrinovic}},
  \bibinfo {author} {\bibfnamefont {J.}~\bibnamefont {Huang}}, \bibinfo
  {author} {\bibfnamefont {M.}~\bibnamefont {Baryakhtar}}, \ and\ \bibinfo
  {author} {\bibfnamefont {W.~E.}\ \bibnamefont {East}},\ }\href {\doibase
  10.1103/PhysRevD.107.075025} {\bibfield  {journal} {\bibinfo  {journal}
  {Phys. Rev. D}\ }\textbf {\bibinfo {volume} {107}},\ \bibinfo {pages}
  {075025} (\bibinfo {year} {2023})},\ \Eprint
  {http://arxiv.org/abs/2212.09772} {arXiv:2212.09772 [astro-ph.HE]}
  \BibitemShut {NoStop}%
\bibitem [{\citenamefont {Spieksma}\ \emph {et~al.}(2023)\citenamefont
  {Spieksma}, \citenamefont {Cannizzaro}, \citenamefont {Ikeda}, \citenamefont
  {Cardoso},\ and\ \citenamefont {Chen}}]{Spieksma:2023vwl}%
  \BibitemOpen
  \bibfield  {author} {\bibinfo {author} {\bibfnamefont {T.~F.~M.}\
  \bibnamefont {Spieksma}}, \bibinfo {author} {\bibfnamefont {E.}~\bibnamefont
  {Cannizzaro}}, \bibinfo {author} {\bibfnamefont {T.}~\bibnamefont {Ikeda}},
  \bibinfo {author} {\bibfnamefont {V.}~\bibnamefont {Cardoso}}, \ and\
  \bibinfo {author} {\bibfnamefont {Y.}~\bibnamefont {Chen}},\ }\href {\doibase
  10.1103/PhysRevD.108.063013} {\bibfield  {journal} {\bibinfo  {journal}
  {Phys. Rev. D}\ }\textbf {\bibinfo {volume} {108}},\ \bibinfo {pages}
  {063013} (\bibinfo {year} {2023})},\ \Eprint
  {http://arxiv.org/abs/2306.16447} {arXiv:2306.16447 [gr-qc]} \BibitemShut
  {NoStop}%
\bibitem [{\citenamefont {Chakraborty}\ \emph {et~al.}(2024)\citenamefont
  {Chakraborty}, \citenamefont {Patil},\ and\ \citenamefont
  {Akash}}]{Chakraborty:2024aug}%
  \BibitemOpen
  \bibfield  {author} {\bibinfo {author} {\bibfnamefont {C.}~\bibnamefont
  {Chakraborty}}, \bibinfo {author} {\bibfnamefont {P.}~\bibnamefont {Patil}},
  \ and\ \bibinfo {author} {\bibfnamefont {G.}~\bibnamefont {Akash}},\ }\href
  {\doibase 10.1103/PhysRevD.109.064062} {\bibfield  {journal} {\bibinfo
  {journal} {Phys. Rev. D}\ }\textbf {\bibinfo {volume} {109}},\ \bibinfo
  {pages} {064062} (\bibinfo {year} {2024})},\ \Eprint
  {http://arxiv.org/abs/2401.13347} {arXiv:2401.13347 [astro-ph.HE]}
  \BibitemShut {NoStop}%
\bibitem [{\citenamefont {Alexander}\ \emph {et~al.}(2023)\citenamefont
  {Alexander}, \citenamefont {Gabadadze}, \citenamefont {Jenks},\ and\
  \citenamefont {Yunes}}]{Alexander:2022avt}%
  \BibitemOpen
  \bibfield  {author} {\bibinfo {author} {\bibfnamefont {S.}~\bibnamefont
  {Alexander}}, \bibinfo {author} {\bibfnamefont {G.}~\bibnamefont
  {Gabadadze}}, \bibinfo {author} {\bibfnamefont {L.}~\bibnamefont {Jenks}}, \
  and\ \bibinfo {author} {\bibfnamefont {N.}~\bibnamefont {Yunes}},\ }\href
  {\doibase 10.1103/PhysRevD.107.084016} {\bibfield  {journal} {\bibinfo
  {journal} {Phys. Rev. D}\ }\textbf {\bibinfo {volume} {107}},\ \bibinfo
  {pages} {084016} (\bibinfo {year} {2023})},\ \Eprint
  {http://arxiv.org/abs/2201.02220} {arXiv:2201.02220 [gr-qc]} \BibitemShut
  {NoStop}%
\bibitem [{\citenamefont {Richards}\ \emph {et~al.}(2023)\citenamefont
  {Richards}, \citenamefont {Dima},\ and\ \citenamefont
  {Witek}}]{Richards:2023xsr}%
  \BibitemOpen
  \bibfield  {author} {\bibinfo {author} {\bibfnamefont {C.}~\bibnamefont
  {Richards}}, \bibinfo {author} {\bibfnamefont {A.}~\bibnamefont {Dima}}, \
  and\ \bibinfo {author} {\bibfnamefont {H.}~\bibnamefont {Witek}},\ }\href
  {\doibase 10.1103/PhysRevD.108.044078} {\bibfield  {journal} {\bibinfo
  {journal} {Phys. Rev. D}\ }\textbf {\bibinfo {volume} {108}},\ \bibinfo
  {pages} {044078} (\bibinfo {year} {2023})},\ \Eprint
  {http://arxiv.org/abs/2305.07704} {arXiv:2305.07704 [gr-qc]} \BibitemShut
  {NoStop}%
\bibitem [{\citenamefont {Preskill}(1984)}]{Preskill:1984gd}%
  \BibitemOpen
  \bibfield  {author} {\bibinfo {author} {\bibfnamefont {J.}~\bibnamefont
  {Preskill}},\ }\href {\doibase 10.1146/annurev.ns.34.120184.002333}
  {\bibfield  {journal} {\bibinfo  {journal} {Ann. Rev. Nucl. Part. Sci.}\
  }\textbf {\bibinfo {volume} {34}},\ \bibinfo {pages} {461} (\bibinfo {year}
  {1984})}\BibitemShut {NoStop}%
\bibitem [{\citenamefont {Thomson}(2009)}]{thomson_2009}%
  \BibitemOpen
  \bibfield  {author} {\bibinfo {author} {\bibfnamefont {J.~J.}\ \bibnamefont
  {Thomson}},\ }\href {\doibase 10.1017/CBO9780511694141} {\emph {\bibinfo
  {title} {Elements of the Mathematical Theory of Electricity and
  Magnetism}}},\ \bibinfo {edition} {4th}\ ed.,\ Cambridge Library Collection -
  Mathematics\ (\bibinfo  {publisher} {Cambridge University Press},\ \bibinfo
  {year} {2009})\BibitemShut {NoStop}%
\bibitem [{\citenamefont {Dirac}(1931)}]{Dirac:1931kp}%
  \BibitemOpen
  \bibfield  {author} {\bibinfo {author} {\bibfnamefont {P.~A.~M.}\
  \bibnamefont {Dirac}},\ }\href {\doibase 10.1098/rspa.1931.0130} {\bibfield
  {journal} {\bibinfo  {journal} {Proc. Roy. Soc. Lond. A}\ }\textbf {\bibinfo
  {volume} {133}},\ \bibinfo {pages} {60} (\bibinfo {year} {1931})}\BibitemShut
  {NoStop}%
\bibitem [{\citenamefont {'t~Hooft}(1974)}]{tHooft:1974kcl}%
  \BibitemOpen
  \bibfield  {author} {\bibinfo {author} {\bibfnamefont {G.}~\bibnamefont
  {'t~Hooft}},\ }\href {\doibase 10.1016/0550-3213(74)90486-6} {\bibfield
  {journal} {\bibinfo  {journal} {Nucl. Phys. B}\ }\textbf {\bibinfo {volume}
  {79}},\ \bibinfo {pages} {276} (\bibinfo {year} {1974})}\BibitemShut
  {NoStop}%
\bibitem [{\citenamefont {Polyakov}(1974)}]{Polyakov:1974ek}%
  \BibitemOpen
  \bibfield  {author} {\bibinfo {author} {\bibfnamefont {A.~M.}\ \bibnamefont
  {Polyakov}},\ }\href@noop {} {\bibfield  {journal} {\bibinfo  {journal} {JETP
  Lett.}\ }\textbf {\bibinfo {volume} {20}},\ \bibinfo {pages} {194} (\bibinfo
  {year} {1974})}\BibitemShut {NoStop}%
\bibitem [{\citenamefont {Ortin}(2015)}]{Ortin:2015hya}%
  \BibitemOpen
  \bibfield  {author} {\bibinfo {author} {\bibfnamefont {T.}~\bibnamefont
  {Ortin}},\ }\href {\doibase 10.1017/CBO9781139019750} {\emph {\bibinfo
  {title} {{Gravity and Strings}}}},\ \bibinfo {edition} {2nd}\ ed.,\ Cambridge
  Monographs on Mathematical Physics\ (\bibinfo  {publisher} {Cambridge
  University Press},\ \bibinfo {year} {2015})\BibitemShut {NoStop}%
\bibitem [{\citenamefont {Giacomelli}\ and\ \citenamefont
  {Patrizii}(2003)}]{Giacomelli:2003yu}%
  \BibitemOpen
  \bibfield  {author} {\bibinfo {author} {\bibfnamefont {G.}~\bibnamefont
  {Giacomelli}}\ and\ \bibinfo {author} {\bibfnamefont {L.}~\bibnamefont
  {Patrizii}},\ }\href@noop {} {\bibfield  {journal} {\bibinfo  {journal} {ICTP
  Lect. Notes Ser.}\ }\textbf {\bibinfo {volume} {14}},\ \bibinfo {pages} {121}
  (\bibinfo {year} {2003})},\ \Eprint {http://arxiv.org/abs/hep-ex/0302011}
  {arXiv:hep-ex/0302011} \BibitemShut {NoStop}%
\bibitem [{\citenamefont {Aad}\ \emph {et~al.}(2023)\citenamefont {Aad} \emph
  {et~al.}}]{ATLAS:2023esy}%
  \BibitemOpen
  \bibfield  {author} {\bibinfo {author} {\bibfnamefont {G.}~\bibnamefont
  {Aad}} \emph {et~al.} (\bibinfo {collaboration} {ATLAS}),\ }\href {\doibase
  10.1007/JHEP11(2023)112} {\bibfield  {journal} {\bibinfo  {journal} {JHEP}\
  }\textbf {\bibinfo {volume} {11}},\ \bibinfo {pages} {112} (\bibinfo {year}
  {2023})},\ \Eprint {http://arxiv.org/abs/2308.04835} {arXiv:2308.04835
  [hep-ex]} \BibitemShut {NoStop}%
\bibitem [{\citenamefont {Workman}\ and\ \citenamefont
  {Others}(2022)}]{Workman:2022ynf}%
  \BibitemOpen
  \bibfield  {author} {\bibinfo {author} {\bibfnamefont {R.~L.}\ \bibnamefont
  {Workman}}\ and\ \bibinfo {author} {\bibnamefont {Others}} (\bibinfo
  {collaboration} {Particle Data Group}),\ }\href {\doibase
  10.1093/ptep/ptac097} {\bibfield  {journal} {\bibinfo  {journal} {PTEP}\
  }\textbf {\bibinfo {volume} {2022}},\ \bibinfo {pages} {083C01} (\bibinfo
  {year} {2022})}\BibitemShut {NoStop}%
\bibitem [{\citenamefont {Gibbons}(1991)}]{Gibbons:1990um}%
  \BibitemOpen
  \bibfield  {author} {\bibinfo {author} {\bibfnamefont {G.~W.}\ \bibnamefont
  {Gibbons}},\ }\href {\doibase 10.1007/3-540-54293-0_24} {\bibfield  {journal}
  {\bibinfo  {journal} {Lect. Notes Phys.}\ }\textbf {\bibinfo {volume}
  {383}},\ \bibinfo {pages} {110} (\bibinfo {year} {1991})},\ \Eprint
  {http://arxiv.org/abs/1109.3538} {arXiv:1109.3538 [gr-qc]} \BibitemShut
  {NoStop}%
\bibitem [{\citenamefont {Ortiz}(1992)}]{Ortiz:1991eu}%
  \BibitemOpen
  \bibfield  {author} {\bibinfo {author} {\bibfnamefont {M.~E.}\ \bibnamefont
  {Ortiz}},\ }\href {\doibase 10.1103/PhysRevD.45.R2586} {\bibfield  {journal}
  {\bibinfo  {journal} {Phys. Rev. D}\ }\textbf {\bibinfo {volume} {45}},\
  \bibinfo {pages} {R2586} (\bibinfo {year} {1992})}\BibitemShut {NoStop}%
\bibitem [{\citenamefont {Lee}\ \emph {et~al.}(1992{\natexlab{a}})\citenamefont
  {Lee}, \citenamefont {Nair},\ and\ \citenamefont {Weinberg}}]{Lee:1991vy}%
  \BibitemOpen
  \bibfield  {author} {\bibinfo {author} {\bibfnamefont {K.-M.}\ \bibnamefont
  {Lee}}, \bibinfo {author} {\bibfnamefont {V.~P.}\ \bibnamefont {Nair}}, \
  and\ \bibinfo {author} {\bibfnamefont {E.~J.}\ \bibnamefont {Weinberg}},\
  }\href {\doibase 10.1103/PhysRevD.45.2751} {\bibfield  {journal} {\bibinfo
  {journal} {Phys. Rev. D}\ }\textbf {\bibinfo {volume} {45}},\ \bibinfo
  {pages} {2751} (\bibinfo {year} {1992}{\natexlab{a}})},\ \Eprint
  {http://arxiv.org/abs/hep-th/9112008} {arXiv:hep-th/9112008} \BibitemShut
  {NoStop}%
\bibitem [{\citenamefont {Lee}\ \emph {et~al.}(1992{\natexlab{b}})\citenamefont
  {Lee}, \citenamefont {Nair},\ and\ \citenamefont {Weinberg}}]{Lee:1991qs}%
  \BibitemOpen
  \bibfield  {author} {\bibinfo {author} {\bibfnamefont {K.-M.}\ \bibnamefont
  {Lee}}, \bibinfo {author} {\bibfnamefont {V.~P.}\ \bibnamefont {Nair}}, \
  and\ \bibinfo {author} {\bibfnamefont {E.~J.}\ \bibnamefont {Weinberg}},\
  }\href {\doibase 10.1103/PhysRevLett.68.1100} {\bibfield  {journal} {\bibinfo
   {journal} {Phys. Rev. Lett.}\ }\textbf {\bibinfo {volume} {68}},\ \bibinfo
  {pages} {1100} (\bibinfo {year} {1992}{\natexlab{b}})},\ \Eprint
  {http://arxiv.org/abs/hep-th/9111045} {arXiv:hep-th/9111045} \BibitemShut
  {NoStop}%
\bibitem [{\citenamefont {Stojkovic}\ and\ \citenamefont
  {Freese}(2005)}]{Stojkovic:2004hz}%
  \BibitemOpen
  \bibfield  {author} {\bibinfo {author} {\bibfnamefont {D.}~\bibnamefont
  {Stojkovic}}\ and\ \bibinfo {author} {\bibfnamefont {K.}~\bibnamefont
  {Freese}},\ }\href {\doibase 10.1016/j.physletb.2004.12.019} {\bibfield
  {journal} {\bibinfo  {journal} {Phys. Lett. B}\ }\textbf {\bibinfo {volume}
  {606}},\ \bibinfo {pages} {251} (\bibinfo {year} {2005})},\ \Eprint
  {http://arxiv.org/abs/hep-ph/0403248} {arXiv:hep-ph/0403248} \BibitemShut
  {NoStop}%
\bibitem [{\citenamefont {Gibbons}(1977)}]{Gibbons:1976sm}%
  \BibitemOpen
  \bibfield  {author} {\bibinfo {author} {\bibfnamefont {G.~W.}\ \bibnamefont
  {Gibbons}},\ }\href {\doibase 10.1103/PhysRevD.15.3530} {\bibfield  {journal}
  {\bibinfo  {journal} {Phys. Rev. D}\ }\textbf {\bibinfo {volume} {15}},\
  \bibinfo {pages} {3530} (\bibinfo {year} {1977})}\BibitemShut {NoStop}%
\bibitem [{\citenamefont {Profumo}(2024)}]{Profumo:2024fxq}%
  \BibitemOpen
  \bibfield  {author} {\bibinfo {author} {\bibfnamefont {S.}~\bibnamefont
  {Profumo}},\ }\href@noop {} {\  (\bibinfo {year} {2024})},\ \Eprint
  {http://arxiv.org/abs/2405.00546} {arXiv:2405.00546 [astro-ph.HE]}
  \BibitemShut {NoStop}%
\bibitem [{\citenamefont {Gibbons}(1975)}]{Gibbons:1975kk}%
  \BibitemOpen
  \bibfield  {author} {\bibinfo {author} {\bibfnamefont {G.~W.}\ \bibnamefont
  {Gibbons}},\ }\href {\doibase 10.1007/BF01609829} {\bibfield  {journal}
  {\bibinfo  {journal} {Commun. Math. Phys.}\ }\textbf {\bibinfo {volume}
  {44}},\ \bibinfo {pages} {245} (\bibinfo {year} {1975})}\BibitemShut
  {NoStop}%
\bibitem [{\citenamefont {Maldacena}(2021)}]{Maldacena:2020skw}%
  \BibitemOpen
  \bibfield  {author} {\bibinfo {author} {\bibfnamefont {J.}~\bibnamefont
  {Maldacena}},\ }\href {\doibase 10.1007/JHEP04(2021)079} {\bibfield
  {journal} {\bibinfo  {journal} {JHEP}\ }\textbf {\bibinfo {volume} {04}},\
  \bibinfo {pages} {079} (\bibinfo {year} {2021})},\ \Eprint
  {http://arxiv.org/abs/2004.06084} {arXiv:2004.06084 [hep-th]} \BibitemShut
  {NoStop}%
\bibitem [{\citenamefont {Liu}\ \emph {et~al.}(2020)\citenamefont {Liu},
  \citenamefont {Christiansen}, \citenamefont {Guo}, \citenamefont {Cai},\ and\
  \citenamefont {Kim}}]{Liu:2020vsy}%
  \BibitemOpen
  \bibfield  {author} {\bibinfo {author} {\bibfnamefont {L.}~\bibnamefont
  {Liu}}, \bibinfo {author} {\bibfnamefont {O.}~\bibnamefont {Christiansen}},
  \bibinfo {author} {\bibfnamefont {Z.-K.}\ \bibnamefont {Guo}}, \bibinfo
  {author} {\bibfnamefont {R.-G.}\ \bibnamefont {Cai}}, \ and\ \bibinfo
  {author} {\bibfnamefont {S.~P.}\ \bibnamefont {Kim}},\ }\href {\doibase
  10.1103/PhysRevD.102.103520} {\bibfield  {journal} {\bibinfo  {journal}
  {Phys. Rev. D}\ }\textbf {\bibinfo {volume} {102}},\ \bibinfo {pages}
  {103520} (\bibinfo {year} {2020})},\ \Eprint
  {http://arxiv.org/abs/2008.02326} {arXiv:2008.02326 [gr-qc]} \BibitemShut
  {NoStop}%
\bibitem [{\citenamefont {Liu}\ \emph {et~al.}(2021)\citenamefont {Liu},
  \citenamefont {Christiansen}, \citenamefont {Ruan}, \citenamefont {Guo},
  \citenamefont {Cai},\ and\ \citenamefont {Kim}}]{Liu:2020bag}%
  \BibitemOpen
  \bibfield  {author} {\bibinfo {author} {\bibfnamefont {L.}~\bibnamefont
  {Liu}}, \bibinfo {author} {\bibfnamefont {O.}~\bibnamefont {Christiansen}},
  \bibinfo {author} {\bibfnamefont {W.-H.}\ \bibnamefont {Ruan}}, \bibinfo
  {author} {\bibfnamefont {Z.-K.}\ \bibnamefont {Guo}}, \bibinfo {author}
  {\bibfnamefont {R.-G.}\ \bibnamefont {Cai}}, \ and\ \bibinfo {author}
  {\bibfnamefont {S.~P.}\ \bibnamefont {Kim}},\ }\href {\doibase
  10.1140/epjc/s10052-021-09849-4} {\bibfield  {journal} {\bibinfo  {journal}
  {Eur. Phys. J. C}\ }\textbf {\bibinfo {volume} {81}},\ \bibinfo {pages}
  {1048} (\bibinfo {year} {2021})},\ \Eprint {http://arxiv.org/abs/2011.13586}
  {arXiv:2011.13586 [gr-qc]} \BibitemShut {NoStop}%
\bibitem [{\citenamefont {Carullo}\ \emph {et~al.}(2022)\citenamefont
  {Carullo}, \citenamefont {Laghi}, \citenamefont {Johnson-McDaniel},
  \citenamefont {Del~Pozzo}, \citenamefont {Dias}, \citenamefont {Godazgar},\
  and\ \citenamefont {Santos}}]{Carullo:2021oxn}%
  \BibitemOpen
  \bibfield  {author} {\bibinfo {author} {\bibfnamefont {G.}~\bibnamefont
  {Carullo}}, \bibinfo {author} {\bibfnamefont {D.}~\bibnamefont {Laghi}},
  \bibinfo {author} {\bibfnamefont {N.~K.}\ \bibnamefont {Johnson-McDaniel}},
  \bibinfo {author} {\bibfnamefont {W.}~\bibnamefont {Del~Pozzo}}, \bibinfo
  {author} {\bibfnamefont {O.~J.~C.}\ \bibnamefont {Dias}}, \bibinfo {author}
  {\bibfnamefont {M.}~\bibnamefont {Godazgar}}, \ and\ \bibinfo {author}
  {\bibfnamefont {J.~E.}\ \bibnamefont {Santos}},\ }\href {\doibase
  10.1103/PhysRevD.105.062009} {\bibfield  {journal} {\bibinfo  {journal}
  {Phys. Rev. D}\ }\textbf {\bibinfo {volume} {105}},\ \bibinfo {pages}
  {062009} (\bibinfo {year} {2022})},\ \Eprint
  {http://arxiv.org/abs/2109.13961} {arXiv:2109.13961 [gr-qc]} \BibitemShut
  {NoStop}%
\bibitem [{\citenamefont {Chen}\ \emph {et~al.}(2023)\citenamefont {Chen},
  \citenamefont {Kim},\ and\ \citenamefont {Liu}}]{Chen:2022qvg}%
  \BibitemOpen
  \bibfield  {author} {\bibinfo {author} {\bibfnamefont {Z.-C.}\ \bibnamefont
  {Chen}}, \bibinfo {author} {\bibfnamefont {S.~P.}\ \bibnamefont {Kim}}, \
  and\ \bibinfo {author} {\bibfnamefont {L.}~\bibnamefont {Liu}},\ }\href
  {\doibase 10.1088/1572-9494/acce98} {\bibfield  {journal} {\bibinfo
  {journal} {Commun. Theor. Phys.}\ }\textbf {\bibinfo {volume} {75}},\
  \bibinfo {pages} {065401} (\bibinfo {year} {2023})},\ \Eprint
  {http://arxiv.org/abs/2210.15564} {arXiv:2210.15564 [gr-qc]} \BibitemShut
  {NoStop}%
\bibitem [{\citenamefont {Pere\~niguez}(2023)}]{Pereniguez:2023wxf}%
  \BibitemOpen
  \bibfield  {author} {\bibinfo {author} {\bibfnamefont {D.}~\bibnamefont
  {Pere\~niguez}},\ }\href {\doibase 10.1103/PhysRevD.108.084046} {\bibfield
  {journal} {\bibinfo  {journal} {Phys. Rev. D}\ }\textbf {\bibinfo {volume}
  {108}},\ \bibinfo {pages} {084046} (\bibinfo {year} {2023})},\ \Eprint
  {http://arxiv.org/abs/2302.10942} {arXiv:2302.10942 [gr-qc]} \BibitemShut
  {NoStop}%
\bibitem [{\citenamefont {De~Felice}\ and\ \citenamefont
  {Tsujikawa}(2024{\natexlab{a}})}]{DeFelice:2023rra}%
  \BibitemOpen
  \bibfield  {author} {\bibinfo {author} {\bibfnamefont {A.}~\bibnamefont
  {De~Felice}}\ and\ \bibinfo {author} {\bibfnamefont {S.}~\bibnamefont
  {Tsujikawa}},\ }\href {\doibase 10.1103/PhysRevD.109.084022} {\bibfield
  {journal} {\bibinfo  {journal} {Phys. Rev. D}\ }\textbf {\bibinfo {volume}
  {109}},\ \bibinfo {pages} {084022} (\bibinfo {year} {2024}{\natexlab{a}})},\
  \Eprint {http://arxiv.org/abs/2312.03191} {arXiv:2312.03191 [gr-qc]}
  \BibitemShut {NoStop}%
\bibitem [{\citenamefont {Bai}\ and\ \citenamefont
  {Orlofsky}(2020)}]{Bai:2019zcd}%
  \BibitemOpen
  \bibfield  {author} {\bibinfo {author} {\bibfnamefont {Y.}~\bibnamefont
  {Bai}}\ and\ \bibinfo {author} {\bibfnamefont {N.}~\bibnamefont {Orlofsky}},\
  }\href {\doibase 10.1103/PhysRevD.101.055006} {\bibfield  {journal} {\bibinfo
   {journal} {Phys. Rev. D}\ }\textbf {\bibinfo {volume} {101}},\ \bibinfo
  {pages} {055006} (\bibinfo {year} {2020})},\ \Eprint
  {http://arxiv.org/abs/1906.04858} {arXiv:1906.04858 [hep-ph]} \BibitemShut
  {NoStop}%
\bibitem [{\citenamefont {Kritos}\ and\ \citenamefont
  {Silk}(2022)}]{Kritos:2021nsf}%
  \BibitemOpen
  \bibfield  {author} {\bibinfo {author} {\bibfnamefont {K.}~\bibnamefont
  {Kritos}}\ and\ \bibinfo {author} {\bibfnamefont {J.}~\bibnamefont {Silk}},\
  }\href {\doibase 10.1103/PhysRevD.105.063011} {\bibfield  {journal} {\bibinfo
   {journal} {Phys. Rev. D}\ }\textbf {\bibinfo {volume} {105}},\ \bibinfo
  {pages} {063011} (\bibinfo {year} {2022})},\ \Eprint
  {http://arxiv.org/abs/2109.09769} {arXiv:2109.09769 [gr-qc]} \BibitemShut
  {NoStop}%
\bibitem [{\citenamefont {Turner}\ \emph {et~al.}(1982)\citenamefont {Turner},
  \citenamefont {Parker},\ and\ \citenamefont {Bogdan}}]{Turner:1982ag}%
  \BibitemOpen
  \bibfield  {author} {\bibinfo {author} {\bibfnamefont {M.~S.}\ \bibnamefont
  {Turner}}, \bibinfo {author} {\bibfnamefont {E.~N.}\ \bibnamefont {Parker}},
  \ and\ \bibinfo {author} {\bibfnamefont {T.~J.}\ \bibnamefont {Bogdan}},\
  }\href {\doibase 10.1103/PhysRevD.26.1296} {\bibfield  {journal} {\bibinfo
  {journal} {Phys. Rev. D}\ }\textbf {\bibinfo {volume} {26}},\ \bibinfo
  {pages} {1296} (\bibinfo {year} {1982})}\BibitemShut {NoStop}%
\bibitem [{\citenamefont {Kobayashi}\ and\ \citenamefont
  {Perri}(2023)}]{Kobayashi:2023ryr}%
  \BibitemOpen
  \bibfield  {author} {\bibinfo {author} {\bibfnamefont {T.}~\bibnamefont
  {Kobayashi}}\ and\ \bibinfo {author} {\bibfnamefont {D.}~\bibnamefont
  {Perri}},\ }\href {\doibase 10.1103/PhysRevD.108.083005} {\bibfield
  {journal} {\bibinfo  {journal} {Phys. Rev. D}\ }\textbf {\bibinfo {volume}
  {108}},\ \bibinfo {pages} {083005} (\bibinfo {year} {2023})},\ \Eprint
  {http://arxiv.org/abs/2307.07553} {arXiv:2307.07553 [hep-ph]} \BibitemShut
  {NoStop}%
\bibitem [{\citenamefont {Zhang}\ and\ \citenamefont
  {Zhang}(2024)}]{Zhang:2023zmb}%
  \BibitemOpen
  \bibfield  {author} {\bibinfo {author} {\bibfnamefont {C.}~\bibnamefont
  {Zhang}}\ and\ \bibinfo {author} {\bibfnamefont {X.}~\bibnamefont {Zhang}},\
  }\href {\doibase 10.1140/epjc/s10052-024-12383-8} {\bibfield  {journal}
  {\bibinfo  {journal} {Eur. Phys. J. C}\ }\textbf {\bibinfo {volume} {84}},\
  \bibinfo {pages} {100} (\bibinfo {year} {2024})},\ \Eprint
  {http://arxiv.org/abs/2308.07166} {arXiv:2308.07166 [hep-ph]} \BibitemShut
  {NoStop}%
\bibitem [{\citenamefont {Wang}\ and\ \citenamefont
  {Deng}(2024)}]{Wang:2023qxj}%
  \BibitemOpen
  \bibfield  {author} {\bibinfo {author} {\bibfnamefont {X.-Z.}\ \bibnamefont
  {Wang}}\ and\ \bibinfo {author} {\bibfnamefont {C.-M.}\ \bibnamefont
  {Deng}},\ }\href {\doibase 10.1140/epjc/s10052-024-12387-4} {\bibfield
  {journal} {\bibinfo  {journal} {Eur. Phys. J. C}\ }\textbf {\bibinfo {volume}
  {84}},\ \bibinfo {pages} {31} (\bibinfo {year} {2024})},\ \Eprint
  {http://arxiv.org/abs/2401.00555} {arXiv:2401.00555 [astro-ph.HE]}
  \BibitemShut {NoStop}%
\bibitem [{\citenamefont {Carr}\ \emph {et~al.}(2021)\citenamefont {Carr},
  \citenamefont {Kohri}, \citenamefont {Sendouda},\ and\ \citenamefont
  {Yokoyama}}]{Carr:2020gox}%
  \BibitemOpen
  \bibfield  {author} {\bibinfo {author} {\bibfnamefont {B.}~\bibnamefont
  {Carr}}, \bibinfo {author} {\bibfnamefont {K.}~\bibnamefont {Kohri}},
  \bibinfo {author} {\bibfnamefont {Y.}~\bibnamefont {Sendouda}}, \ and\
  \bibinfo {author} {\bibfnamefont {J.}~\bibnamefont {Yokoyama}},\ }\href
  {\doibase 10.1088/1361-6633/ac1e31} {\bibfield  {journal} {\bibinfo
  {journal} {Rept. Prog. Phys.}\ }\textbf {\bibinfo {volume} {84}},\ \bibinfo
  {pages} {116902} (\bibinfo {year} {2021})},\ \Eprint
  {http://arxiv.org/abs/2002.12778} {arXiv:2002.12778 [astro-ph.CO]}
  \BibitemShut {NoStop}%
\bibitem [{\citenamefont {Escriv\`a}\ \emph {et~al.}(2022)\citenamefont
  {Escriv\`a}, \citenamefont {Kuhnel},\ and\ \citenamefont
  {Tada}}]{Escriva:2022duf}%
  \BibitemOpen
  \bibfield  {author} {\bibinfo {author} {\bibfnamefont {A.}~\bibnamefont
  {Escriv\`a}}, \bibinfo {author} {\bibfnamefont {F.}~\bibnamefont {Kuhnel}}, \
  and\ \bibinfo {author} {\bibfnamefont {Y.}~\bibnamefont {Tada}},\ }\href
  {\doibase 10.1016/B978-0-32-395636-9.00012-8} {\  (\bibinfo {year} {2022}),\
  10.1016/B978-0-32-395636-9.00012-8},\ \Eprint
  {http://arxiv.org/abs/2211.05767} {arXiv:2211.05767 [astro-ph.CO]}
  \BibitemShut {NoStop}%
\bibitem [{\citenamefont {Dolan}(2007)}]{Dolan:2007mj}%
  \BibitemOpen
  \bibfield  {author} {\bibinfo {author} {\bibfnamefont {S.~R.}\ \bibnamefont
  {Dolan}},\ }\href {\doibase 10.1103/PhysRevD.76.084001} {\bibfield  {journal}
  {\bibinfo  {journal} {Phys. Rev. D}\ }\textbf {\bibinfo {volume} {76}},\
  \bibinfo {pages} {084001} (\bibinfo {year} {2007})},\ \Eprint
  {http://arxiv.org/abs/0705.2880} {arXiv:0705.2880 [gr-qc]} \BibitemShut
  {NoStop}%
\bibitem [{\citenamefont {Wu}\ and\ \citenamefont {Yang}(1976)}]{Wu:1976ge}%
  \BibitemOpen
  \bibfield  {author} {\bibinfo {author} {\bibfnamefont {T.~T.}\ \bibnamefont
  {Wu}}\ and\ \bibinfo {author} {\bibfnamefont {C.~N.}\ \bibnamefont {Yang}},\
  }\href {\doibase 10.1016/0550-3213(76)90143-7} {\bibfield  {journal}
  {\bibinfo  {journal} {Nucl. Phys. B}\ }\textbf {\bibinfo {volume} {107}},\
  \bibinfo {pages} {365} (\bibinfo {year} {1976})}\BibitemShut {NoStop}%
\bibitem [{\citenamefont {Konoplya}\ and\ \citenamefont
  {Zhidenko}(2013)}]{Konoplya:2013rxa}%
  \BibitemOpen
  \bibfield  {author} {\bibinfo {author} {\bibfnamefont {R.~A.}\ \bibnamefont
  {Konoplya}}\ and\ \bibinfo {author} {\bibfnamefont {A.}~\bibnamefont
  {Zhidenko}},\ }\href {\doibase 10.1103/PhysRevD.88.024054} {\bibfield
  {journal} {\bibinfo  {journal} {Phys. Rev. D}\ }\textbf {\bibinfo {volume}
  {88}},\ \bibinfo {pages} {024054} (\bibinfo {year} {2013})},\ \Eprint
  {http://arxiv.org/abs/1307.1812} {arXiv:1307.1812 [gr-qc]} \BibitemShut
  {NoStop}%
\bibitem [{\citenamefont {Arvanitaki}\ and\ \citenamefont
  {Dubovsky}(2011)}]{Arvanitaki:2010sy}%
  \BibitemOpen
  \bibfield  {author} {\bibinfo {author} {\bibfnamefont {A.}~\bibnamefont
  {Arvanitaki}}\ and\ \bibinfo {author} {\bibfnamefont {S.}~\bibnamefont
  {Dubovsky}},\ }\href {\doibase 10.1103/PhysRevD.83.044026} {\bibfield
  {journal} {\bibinfo  {journal} {Phys. Rev. D}\ }\textbf {\bibinfo {volume}
  {83}},\ \bibinfo {pages} {044026} (\bibinfo {year} {2011})},\ \Eprint
  {http://arxiv.org/abs/1004.3558} {arXiv:1004.3558 [hep-th]} \BibitemShut
  {NoStop}%
\bibitem [{\citenamefont {East}(2022)}]{East:2022ppo}%
  \BibitemOpen
  \bibfield  {author} {\bibinfo {author} {\bibfnamefont {W.~E.}\ \bibnamefont
  {East}},\ }\href {\doibase 10.1103/PhysRevLett.129.141103} {\bibfield
  {journal} {\bibinfo  {journal} {Phys. Rev. Lett.}\ }\textbf {\bibinfo
  {volume} {129}},\ \bibinfo {pages} {141103} (\bibinfo {year} {2022})},\
  \Eprint {http://arxiv.org/abs/2205.03417} {arXiv:2205.03417 [hep-ph]}
  \BibitemShut {NoStop}%
\bibitem [{\citenamefont {Semiz}(1992)}]{Semiz:1991kh}%
  \BibitemOpen
  \bibfield  {author} {\bibinfo {author} {\bibfnamefont {I.}~\bibnamefont
  {Semiz}},\ }\href {\doibase 10.1103/PhysRevD.45.532} {\bibfield  {journal}
  {\bibinfo  {journal} {Phys. Rev. D}\ }\textbf {\bibinfo {volume} {45}},\
  \bibinfo {pages} {532} (\bibinfo {year} {1992})},\ \bibinfo {note} {[Erratum:
  Phys.Rev.D 47, 5615 (1993)]}\BibitemShut {NoStop}%
\bibitem [{\citenamefont {Dyson}\ and\ \citenamefont
  {Pere\~niguez}(2023)}]{Dyson:2023ujk}%
  \BibitemOpen
  \bibfield  {author} {\bibinfo {author} {\bibfnamefont {C.}~\bibnamefont
  {Dyson}}\ and\ \bibinfo {author} {\bibfnamefont {D.}~\bibnamefont
  {Pere\~niguez}},\ }\href {\doibase 10.1103/PhysRevD.108.084064} {\bibfield
  {journal} {\bibinfo  {journal} {Phys. Rev. D}\ }\textbf {\bibinfo {volume}
  {108}},\ \bibinfo {pages} {084064} (\bibinfo {year} {2023})},\ \Eprint
  {http://arxiv.org/abs/2306.15751} {arXiv:2306.15751 [gr-qc]} \BibitemShut
  {NoStop}%
\bibitem [{\citenamefont {Bekenstein}(1996)}]{Bekenstein:1996pn}%
  \BibitemOpen
  \bibfield  {author} {\bibinfo {author} {\bibfnamefont {J.~D.}\ \bibnamefont
  {Bekenstein}},\ }in\ \href@noop {} {\emph {\bibinfo {booktitle} {{2nd
  International Sakharov Conference on Physics}}}}\ (\bibinfo {year} {1996})\
  pp.\ \bibinfo {pages} {216--219},\ \Eprint
  {http://arxiv.org/abs/gr-qc/9605059} {arXiv:gr-qc/9605059} \BibitemShut
  {NoStop}%
\bibitem [{\citenamefont {Ortin}\ and\ \citenamefont
  {Pere\~niguez}(2022)}]{Ortin:2022uxa}%
  \BibitemOpen
  \bibfield  {author} {\bibinfo {author} {\bibfnamefont {T.}~\bibnamefont
  {Ortin}}\ and\ \bibinfo {author} {\bibfnamefont {D.}~\bibnamefont
  {Pere\~niguez}},\ }\href {\doibase 10.1007/JHEP11(2022)081} {\bibfield
  {journal} {\bibinfo  {journal} {JHEP}\ }\textbf {\bibinfo {volume} {11}},\
  \bibinfo {pages} {081} (\bibinfo {year} {2022})},\ \Eprint
  {http://arxiv.org/abs/2207.12008} {arXiv:2207.12008 [hep-th]} \BibitemShut
  {NoStop}%
\bibitem [{\citenamefont {Leaver}(1985)}]{Leaver:1985ax}%
  \BibitemOpen
  \bibfield  {author} {\bibinfo {author} {\bibfnamefont {E.~W.}\ \bibnamefont
  {Leaver}},\ }\href {\doibase 10.1098/rspa.1985.0119} {\bibfield  {journal}
  {\bibinfo  {journal} {Proc. Roy. Soc. Lond. A}\ }\textbf {\bibinfo {volume}
  {402}},\ \bibinfo {pages} {285} (\bibinfo {year} {1985})}\BibitemShut
  {NoStop}%
\bibitem [{\citenamefont {Cardoso}\ and\ \citenamefont
  {Yoshida}(2005)}]{Cardoso:2005vk}%
  \BibitemOpen
  \bibfield  {author} {\bibinfo {author} {\bibfnamefont {V.}~\bibnamefont
  {Cardoso}}\ and\ \bibinfo {author} {\bibfnamefont {S.}~\bibnamefont
  {Yoshida}},\ }\href {\doibase 10.1088/1126-6708/2005/07/009} {\bibfield
  {journal} {\bibinfo  {journal} {JHEP}\ }\textbf {\bibinfo {volume} {07}},\
  \bibinfo {pages} {009} (\bibinfo {year} {2005})},\ \Eprint
  {http://arxiv.org/abs/hep-th/0502206} {arXiv:hep-th/0502206} \BibitemShut
  {NoStop}%
\bibitem [{\citenamefont {Page}(1976{\natexlab{a}})}]{Page:1976df}%
  \BibitemOpen
  \bibfield  {author} {\bibinfo {author} {\bibfnamefont {D.~N.}\ \bibnamefont
  {Page}},\ }\href {\doibase 10.1103/PhysRevD.13.198} {\bibfield  {journal}
  {\bibinfo  {journal} {Phys. Rev. D}\ }\textbf {\bibinfo {volume} {13}},\
  \bibinfo {pages} {198} (\bibinfo {year} {1976}{\natexlab{a}})}\BibitemShut
  {NoStop}%
\bibitem [{\citenamefont {Page}(1976{\natexlab{b}})}]{Page:1976ki}%
  \BibitemOpen
  \bibfield  {author} {\bibinfo {author} {\bibfnamefont {D.~N.}\ \bibnamefont
  {Page}},\ }\href {\doibase 10.1103/PhysRevD.14.3260} {\bibfield  {journal}
  {\bibinfo  {journal} {Phys. Rev. D}\ }\textbf {\bibinfo {volume} {14}},\
  \bibinfo {pages} {3260} (\bibinfo {year} {1976}{\natexlab{b}})}\BibitemShut
  {NoStop}%
\bibitem [{\citenamefont {Page}(1977)}]{Page:1977um}%
  \BibitemOpen
  \bibfield  {author} {\bibinfo {author} {\bibfnamefont {D.~N.}\ \bibnamefont
  {Page}},\ }\href {\doibase 10.1103/PhysRevD.16.2402} {\bibfield  {journal}
  {\bibinfo  {journal} {Phys. Rev. D}\ }\textbf {\bibinfo {volume} {16}},\
  \bibinfo {pages} {2402} (\bibinfo {year} {1977})}\BibitemShut {NoStop}%
\bibitem [{\citenamefont {Carter}(1973)}]{Carter:1973rla}%
  \BibitemOpen
  \bibfield  {author} {\bibinfo {author} {\bibfnamefont {B.}~\bibnamefont
  {Carter}},\ }in\ \href@noop {} {\emph {\bibinfo {booktitle} {{Les Houches
  Summer School of Theoretical Physics}: {Black Holes}}}}\ (\bibinfo {year}
  {1973})\ pp.\ \bibinfo {pages} {57--214}\BibitemShut {NoStop}%
\bibitem [{\citenamefont {Lozano-Tellechea}\ and\ \citenamefont
  {Ortin}(2000)}]{Lozano-Tellechea:1999lwm}%
  \BibitemOpen
  \bibfield  {author} {\bibinfo {author} {\bibfnamefont {E.}~\bibnamefont
  {Lozano-Tellechea}}\ and\ \bibinfo {author} {\bibfnamefont {T.}~\bibnamefont
  {Ortin}},\ }\href {\doibase 10.1016/S0550-3213(99)00762-2} {\bibfield
  {journal} {\bibinfo  {journal} {Nucl. Phys. B}\ }\textbf {\bibinfo {volume}
  {569}},\ \bibinfo {pages} {435} (\bibinfo {year} {2000})},\ \Eprint
  {http://arxiv.org/abs/hep-th/9910020} {arXiv:hep-th/9910020} \BibitemShut
  {NoStop}%
\bibitem [{\citenamefont {De~Felice}\ and\ \citenamefont
  {Tsujikawa}(2024{\natexlab{b}})}]{DeFelice:2024eoj}%
  \BibitemOpen
  \bibfield  {author} {\bibinfo {author} {\bibfnamefont {A.}~\bibnamefont
  {De~Felice}}\ and\ \bibinfo {author} {\bibfnamefont {S.}~\bibnamefont
  {Tsujikawa}},\ }\href@noop {} {\  (\bibinfo {year} {2024}{\natexlab{b}})},\
  \Eprint {http://arxiv.org/abs/2402.08868} {arXiv:2402.08868 [gr-qc]}
  \BibitemShut {NoStop}%
\bibitem [{\citenamefont {Liu}\ and\ \citenamefont
  {Zahed}(2018)}]{Liu:2017spl}%
  \BibitemOpen
  \bibfield  {author} {\bibinfo {author} {\bibfnamefont {Y.}~\bibnamefont
  {Liu}}\ and\ \bibinfo {author} {\bibfnamefont {I.}~\bibnamefont {Zahed}},\
  }\href {\doibase 10.1103/PhysRevLett.120.032001} {\bibfield  {journal}
  {\bibinfo  {journal} {Phys. Rev. Lett.}\ }\textbf {\bibinfo {volume} {120}},\
  \bibinfo {pages} {032001} (\bibinfo {year} {2018})},\ \Eprint
  {http://arxiv.org/abs/1711.08354} {arXiv:1711.08354 [hep-ph]} \BibitemShut
  {NoStop}%
\bibitem [{\citenamefont {Rice}\ and\ \citenamefont
  {Zhang}(2017)}]{Rice:2017avg}%
  \BibitemOpen
  \bibfield  {author} {\bibinfo {author} {\bibfnamefont {J.~R.}\ \bibnamefont
  {Rice}}\ and\ \bibinfo {author} {\bibfnamefont {B.}~\bibnamefont {Zhang}},\
  }\href {\doibase 10.1016/j.jheap.2017.02.002} {\bibfield  {journal} {\bibinfo
   {journal} {JHEAp}\ }\textbf {\bibinfo {volume} {13-14}},\ \bibinfo {pages}
  {22} (\bibinfo {year} {2017})},\ \Eprint {http://arxiv.org/abs/1702.08069}
  {arXiv:1702.08069 [astro-ph.HE]} \BibitemShut {NoStop}%
\bibitem [{\citenamefont {Franciolini}\ \emph {et~al.}(2022)\citenamefont
  {Franciolini}, \citenamefont {Maharana},\ and\ \citenamefont
  {Muia}}]{Franciolini:2022htd}%
  \BibitemOpen
  \bibfield  {author} {\bibinfo {author} {\bibfnamefont {G.}~\bibnamefont
  {Franciolini}}, \bibinfo {author} {\bibfnamefont {A.}~\bibnamefont
  {Maharana}}, \ and\ \bibinfo {author} {\bibfnamefont {F.}~\bibnamefont
  {Muia}},\ }\href {\doibase 10.1103/PhysRevD.106.103520} {\bibfield  {journal}
  {\bibinfo  {journal} {Phys. Rev. D}\ }\textbf {\bibinfo {volume} {106}},\
  \bibinfo {pages} {103520} (\bibinfo {year} {2022})},\ \Eprint
  {http://arxiv.org/abs/2205.02153} {arXiv:2205.02153 [astro-ph.CO]}
  \BibitemShut {NoStop}%
\bibitem [{\citenamefont {Herdeiro}\ and\ \citenamefont
  {Radu}(2014)}]{Herdeiro:2014goa}%
  \BibitemOpen
  \bibfield  {author} {\bibinfo {author} {\bibfnamefont {C.~A.~R.}\
  \bibnamefont {Herdeiro}}\ and\ \bibinfo {author} {\bibfnamefont
  {E.}~\bibnamefont {Radu}},\ }\href {\doibase 10.1103/PhysRevLett.112.221101}
  {\bibfield  {journal} {\bibinfo  {journal} {Phys. Rev. Lett.}\ }\textbf
  {\bibinfo {volume} {112}},\ \bibinfo {pages} {221101} (\bibinfo {year}
  {2014})},\ \Eprint {http://arxiv.org/abs/1403.2757} {arXiv:1403.2757 [gr-qc]}
  \BibitemShut {NoStop}%
\bibitem [{\citenamefont {Bhattacharjee}\ and\ \citenamefont
  {Sigl}(2000)}]{Bhattacharjee:1999mup}%
  \BibitemOpen
  \bibfield  {author} {\bibinfo {author} {\bibfnamefont {P.}~\bibnamefont
  {Bhattacharjee}}\ and\ \bibinfo {author} {\bibfnamefont {G.}~\bibnamefont
  {Sigl}},\ }\href {\doibase 10.1016/S0370-1573(99)00101-5} {\bibfield
  {journal} {\bibinfo  {journal} {Phys. Rept.}\ }\textbf {\bibinfo {volume}
  {327}},\ \bibinfo {pages} {109} (\bibinfo {year} {2000})},\ \Eprint
  {http://arxiv.org/abs/astro-ph/9811011} {arXiv:astro-ph/9811011} \BibitemShut
  {NoStop}%
\bibitem [{\citenamefont {Abbasi}\ \emph {et~al.}(2023)\citenamefont {Abbasi}
  \emph {et~al.}}]{TelescopeArray:2023sbd}%
  \BibitemOpen
  \bibfield  {author} {\bibinfo {author} {\bibfnamefont {R.~U.}\ \bibnamefont
  {Abbasi}} \emph {et~al.} (\bibinfo {collaboration} {Telescope Array}),\
  }\href {\doibase 10.1126/science.abo5095} {\bibfield  {journal} {\bibinfo
  {journal} {Science}\ }\textbf {\bibinfo {volume} {382}},\ \bibinfo {pages}
  {903} (\bibinfo {year} {2023})},\ \Eprint {http://arxiv.org/abs/2311.14231}
  {arXiv:2311.14231 [astro-ph.HE]} \BibitemShut {NoStop}%
\bibitem [{\citenamefont {Predehl}\ \emph {et~al.}(2020)\citenamefont {Predehl}
  \emph {et~al.}}]{Predehl:2020kyq}%
  \BibitemOpen
  \bibfield  {author} {\bibinfo {author} {\bibfnamefont {P.}~\bibnamefont
  {Predehl}} \emph {et~al.},\ }\href {\doibase 10.1038/s41586-020-2979-0}
  {\bibfield  {journal} {\bibinfo  {journal} {Nature}\ }\textbf {\bibinfo
  {volume} {588}},\ \bibinfo {pages} {227} (\bibinfo {year} {2020})},\ \Eprint
  {http://arxiv.org/abs/2012.05840} {arXiv:2012.05840 [astro-ph.GA]}
  \BibitemShut {NoStop}%
\bibitem [{\citenamefont {Mouland}\ and\ \citenamefont
  {Tong}(2024)}]{Mouland:2024zgk}%
  \BibitemOpen
  \bibfield  {author} {\bibinfo {author} {\bibfnamefont {R.}~\bibnamefont
  {Mouland}}\ and\ \bibinfo {author} {\bibfnamefont {D.}~\bibnamefont {Tong}},\
  }\href@noop {} {\  (\bibinfo {year} {2024})},\ \Eprint
  {http://arxiv.org/abs/2401.01924} {arXiv:2401.01924 [hep-th]} \BibitemShut
  {NoStop}%
\bibitem [{\citenamefont {Elgood}\ \emph {et~al.}(2020)\citenamefont {Elgood},
  \citenamefont {Meessen},\ and\ \citenamefont {Ort\'\i{}n}}]{Elgood:2020svt}%
  \BibitemOpen
  \bibfield  {author} {\bibinfo {author} {\bibfnamefont {Z.}~\bibnamefont
  {Elgood}}, \bibinfo {author} {\bibfnamefont {P.}~\bibnamefont {Meessen}}, \
  and\ \bibinfo {author} {\bibfnamefont {T.}~\bibnamefont {Ort\'\i{}n}},\
  }\href {\doibase 10.1007/JHEP09(2020)026} {\bibfield  {journal} {\bibinfo
  {journal} {JHEP}\ }\textbf {\bibinfo {volume} {09}},\ \bibinfo {pages} {026}
  (\bibinfo {year} {2020})},\ \Eprint {http://arxiv.org/abs/2006.02792}
  {arXiv:2006.02792 [hep-th]} \BibitemShut {NoStop}%
\bibitem [{\citenamefont {Barnich}\ and\ \citenamefont
  {Brandt}(2002)}]{Barnich:2001jy}%
  \BibitemOpen
  \bibfield  {author} {\bibinfo {author} {\bibfnamefont {G.}~\bibnamefont
  {Barnich}}\ and\ \bibinfo {author} {\bibfnamefont {F.}~\bibnamefont
  {Brandt}},\ }\href {\doibase 10.1016/S0550-3213(02)00251-1} {\bibfield
  {journal} {\bibinfo  {journal} {Nucl. Phys. B}\ }\textbf {\bibinfo {volume}
  {633}},\ \bibinfo {pages} {3} (\bibinfo {year} {2002})},\ \Eprint
  {http://arxiv.org/abs/hep-th/0111246} {arXiv:hep-th/0111246} \BibitemShut
  {NoStop}%
\bibitem [{\citenamefont {Barnich}\ and\ \citenamefont
  {Compere}(2008)}]{Barnich:2007bf}%
  \BibitemOpen
  \bibfield  {author} {\bibinfo {author} {\bibfnamefont {G.}~\bibnamefont
  {Barnich}}\ and\ \bibinfo {author} {\bibfnamefont {G.}~\bibnamefont
  {Compere}},\ }\href {\doibase 10.1063/1.2889721} {\bibfield  {journal}
  {\bibinfo  {journal} {J. Math. Phys.}\ }\textbf {\bibinfo {volume} {49}},\
  \bibinfo {pages} {042901} (\bibinfo {year} {2008})},\ \Eprint
  {http://arxiv.org/abs/0708.2378} {arXiv:0708.2378 [gr-qc]} \BibitemShut
  {NoStop}%
\bibitem [{\citenamefont {Iyer}\ and\ \citenamefont
  {Wald}(1994)}]{Iyer:1994ys}%
  \BibitemOpen
  \bibfield  {author} {\bibinfo {author} {\bibfnamefont {V.}~\bibnamefont
  {Iyer}}\ and\ \bibinfo {author} {\bibfnamefont {R.~M.}\ \bibnamefont
  {Wald}},\ }\href {\doibase 10.1103/PhysRevD.50.846} {\bibfield  {journal}
  {\bibinfo  {journal} {Phys. Rev. D}\ }\textbf {\bibinfo {volume} {50}},\
  \bibinfo {pages} {846} (\bibinfo {year} {1994})},\ \Eprint
  {http://arxiv.org/abs/gr-qc/9403028} {arXiv:gr-qc/9403028} \BibitemShut
  {NoStop}%
\bibitem [{\citenamefont {Gao}\ and\ \citenamefont {Wald}(2001)}]{Gao:2001ut}%
  \BibitemOpen
  \bibfield  {author} {\bibinfo {author} {\bibfnamefont {S.}~\bibnamefont
  {Gao}}\ and\ \bibinfo {author} {\bibfnamefont {R.~M.}\ \bibnamefont {Wald}},\
  }\href {\doibase 10.1103/PhysRevD.64.084020} {\bibfield  {journal} {\bibinfo
  {journal} {Phys. Rev. D}\ }\textbf {\bibinfo {volume} {64}},\ \bibinfo
  {pages} {084020} (\bibinfo {year} {2001})},\ \Eprint
  {http://arxiv.org/abs/gr-qc/0106071} {arXiv:gr-qc/0106071} \BibitemShut
  {NoStop}%
\bibitem [{gri()}]{grit_website}%
  \BibitemOpen
  \href@noop {} {\enquote {\bibinfo {title} {{grit repo}},}\ }\bibinfo
  {howpublished}
  {\url{https://centra.tecnico.ulisboa.pt/network/grit/files/}}\BibitemShut
  {NoStop}%
\bibitem [{\citenamefont {Benone}\ and\ \citenamefont
  {Crispino}(2019)}]{Benone:2019all}%
  \BibitemOpen
  \bibfield  {author} {\bibinfo {author} {\bibfnamefont {C.~L.}\ \bibnamefont
  {Benone}}\ and\ \bibinfo {author} {\bibfnamefont {L.~C.~B.}\ \bibnamefont
  {Crispino}},\ }\href {\doibase 10.1103/PhysRevD.99.044009} {\bibfield
  {journal} {\bibinfo  {journal} {Phys. Rev. D}\ }\textbf {\bibinfo {volume}
  {99}},\ \bibinfo {pages} {044009} (\bibinfo {year} {2019})},\ \Eprint
  {http://arxiv.org/abs/1901.05592} {arXiv:1901.05592 [gr-qc]} \BibitemShut
  {NoStop}%
\bibitem [{\citenamefont {Jaramillo}\ \emph {et~al.}(2021)\citenamefont
  {Jaramillo}, \citenamefont {Panosso~Macedo},\ and\ \citenamefont
  {Al~Sheikh}}]{Jaramillo:2020tuu}%
  \BibitemOpen
  \bibfield  {author} {\bibinfo {author} {\bibfnamefont {J.~L.}\ \bibnamefont
  {Jaramillo}}, \bibinfo {author} {\bibfnamefont {R.}~\bibnamefont
  {Panosso~Macedo}}, \ and\ \bibinfo {author} {\bibfnamefont {L.}~\bibnamefont
  {Al~Sheikh}},\ }\href {\doibase 10.1103/PhysRevX.11.031003} {\bibfield
  {journal} {\bibinfo  {journal} {Phys. Rev. X}\ }\textbf {\bibinfo {volume}
  {11}},\ \bibinfo {pages} {031003} (\bibinfo {year} {2021})},\ \Eprint
  {http://arxiv.org/abs/2004.06434} {arXiv:2004.06434 [gr-qc]} \BibitemShut
  {NoStop}%
\bibitem [{\citenamefont {Panosso~Macedo}(2024)}]{PanossoMacedo:2023qzp}%
  \BibitemOpen
  \bibfield  {author} {\bibinfo {author} {\bibfnamefont {R.}~\bibnamefont
  {Panosso~Macedo}},\ }\href {\doibase 10.1098/rsta.2023.0046} {\bibfield
  {journal} {\bibinfo  {journal} {Phil. Trans. Roy. Soc. Lond. A}\ }\textbf
  {\bibinfo {volume} {382}},\ \bibinfo {pages} {20230046} (\bibinfo {year}
  {2024})},\ \Eprint {http://arxiv.org/abs/2307.15735} {arXiv:2307.15735
  [gr-qc]} \BibitemShut {NoStop}%
\end{thebibliography}%
\appendix
\onecolumngrid
\section{Dirac's Monopole and Monopole Spherical Harmonics}\label{A1}

Dirac's monopole with charge $P$ is the following Maxwell vector potential $\bold{A}$ in Euclidean space (minus the origin, technically\footnote{See e.g. \cite{Mouland:2024zgk} for a recent discussion on Dirac monopole $U(1)$-bundles.}) with metric $\bold{d s}^{2}$,
\begin{equation}
    \begin{aligned}\label{DiracMonopole}
        \bold{d s}^{2}=dx^{2}+dy^{2}+dz^{2}\, , \ \ \bold{A}=-P\frac{z}{r}\left(\frac{x dy-ydx}{x^{2}+y^{2}}\right)=-P\cos\theta d\phi\, .
    \end{aligned}
\end{equation}
Scalar fields are sections in the associated $U(1)$-line bundle $\varphi\in\mathbb{C}$, where the gauge-covariant derivative is $ \bold{D}=\boldsymbol{\nabla}+i e \bold{A}$, with $e$ the field's charge, and the $U(1)$ action is
\begin{equation}\label{eq:gauge}
\varphi \mapsto e^{i\alpha(x)}\varphi\, , \ \ \ \ \ \bold{A}\mapsto \bold{A}-\frac{1}{e}\ \boldsymbol{\nabla}\alpha(x)\, ,
\end{equation}
where $\alpha(x)\in\mathbb{R}$ is any real function. We notice that $\bold{A}$ is written in an \textit{intermediate gauge}, which is neither regular at $\cos\theta=+1$ nor at $\cos\theta=-1$, and the gauges  $\bold{A}_{\pm}$ that are regular at each pole are achieved by acting with $\alpha_{\pm}=\mp e P \phi $ (we find it more convenient to work in an intermediate gauge, rather than in the gauges that cover one of the poles as Wu and Yang \cite{Wu:1976ge}, by the same reasons that it is more convenient to work with $(\theta,\phi)$ coordinates on the 2-sphere than with stereographic coordinates). Furthermore, smoothness of this local system as a principal $U(1)$-bundle requires Dirac's quantisation condition, $q\equiv -eP=\pm N/2$ with $N=0,1,2...$. The angular momentum operators introduced in the main text, in spherical coordinates and in the intermediate gauge read
\begin{equation}
    \begin{aligned}
    L_{x}&=i\left(\sin\phi \partial_{\theta}+\cot{\theta}\cos\phi \partial_{\phi} \right)-q \frac{\cos\phi}{\sin \theta}\, ,\ \  L_{y}=i\left(\sin\phi \cot\theta\partial_{\phi}-\cos\phi \partial_{\theta} \right)-q \frac{\sin\phi}{\sin \theta}\, , \ \  L_{z}=-i \partial_{\phi}\, ,\\ 
    L^{2}&=-\partial_{\theta}^{2}-\frac{1}{\sin^{2}\theta}\partial_{\phi}^{2}-\cot\theta \partial_{\theta}-2qi\frac{\cos\theta}{\sin^{2}\theta}\partial_{\phi}+\frac{q^{2}}{\sin^{2}\theta}\, ,
\end{aligned}
\end{equation}
and one can solve explicitly the eigenvalue problem for the spherical monopole harmonics with standard separation of variables and imposing that, in the gauges given by $\alpha_{\pm}$, the solution should be single-valued and regular at $\cos\theta=\pm1$. Then, one finds that the solution in the intermediate gauge reads \cite{Wu:1976ge}
\begin{equation}
\begin{aligned}\label{Yexp}
   & Y_{q,\ell,m}=\Theta_{q,\ell,m}(\theta)e^{i m\phi}=N (1-x)^{\frac{\vert \alpha \vert}{2}}(1+x)^{\frac{\vert \beta \vert}{2}} P^{(\vert \alpha \vert,\vert \beta \vert)}_{\nu}(x)e^{i m\phi}\, , \\
     &  \ell=\lvert q \rvert, \lvert q \rvert+1, ... \, ,  \ \ m=-\ell,-\ell+1,...,\ell\, ,
\end{aligned}
\end{equation}
where $x=\cos\theta$, $P^{(a,b)}_{n}(x)$ denote the usual Jacobi polynomials, the various constants are
\begin{align}
    \alpha=-q-m\, ,\ \ \ \beta=q-m\, , \ \ \  \nu=\ell +m+\frac{\alpha-\lvert \alpha \rvert+\beta-\lvert \beta \rvert}{2} \, ,\ \ \  N=\frac{(-1)^{\frac{\alpha-\lvert \alpha \rvert}{2}}}{\sqrt{4\pi}}\sqrt{\frac{2\ell+1}{ 2^{\lvert \alpha \rvert+\lvert \beta \rvert}}\frac{\nu!(\nu+\lvert \alpha \rvert+\lvert \beta \rvert)!}{(\nu+\lvert \alpha \rvert)!(\nu+\lvert \beta \rvert)!}}\, ,
\end{align}
and $N$ has been fixed so that the usual orthogonality condition holds,
\begin{equation}
    \int_{S^{2}}\bar{Y}_{q,\ell',m'}Y_{q,\ell,m}d\Omega = \delta_{\ell'\ell}\delta_{m'm}\, .
\end{equation}
Table~\ref{monopoleharms} shows the explicit expressions for some low-order monopole harmonics, in the intermediate gauge \eqref{Yexp}. 
\begin{table}[h!]
\centering
\begin{equation*}
\begin{array}{l!{\vline width 1pt}c|c|c}
\sqrt{4\pi}Y_{q,\ell,m} & q=0 & q=1/2 & q=1  \\ 
\Xhline{1pt}
\ell=0 & 1 & - & -  \\ \hline
\ell=1/2 & - & \mp \sqrt{1\mp x} e^{\pm i \frac{1}{2}\phi} & -  \\ \hline
\ell=1 & \begin{aligned} \sqrt{3} x \\ \mp\sqrt{3/2}\sqrt{1-x^{2}}e^{\pm i \phi} \end{aligned} & - & \begin{aligned} -\sqrt{3/2}\sqrt{1-x^{2}} \\ \sqrt{3/4}(1\mp x)e^{\pm i \phi} \end{aligned} \\ \hline
\ell=3/2 & - & \begin{aligned} -\sqrt{1/2}\sqrt{1\mp x}(1\pm 3 x)e^{\pm i\frac{1}{2}\phi} \\ \sqrt{3/2}\sqrt{1\pm x}(1\mp x)e^{\pm i\frac{3}{2}\phi} \end{aligned} & - \\ \Xhline{1pt}
\end{array}
\end{equation*}
\caption{Low-order monopole harmonics in the intermediate gauge. As discussed in the main text, for half-integer values of $q$ monopole harmonics form representations of $\mathfrak{so}(3)$ with half-integer maximum weight $\ell$. For integer $q$'s one has the usual representations, although the modes have different properties with respect to the $q=0$ case, such as the existence of north and south monopole modes.}
\label{monopoleharms}
\end{table}
It is worth paying attention to the behaviour of $Y^{\pm}_{q,\ell,m}$ (the result of acting on $Y_{q,\ell,m}$ in \eqref{Yexp}, written in the intermediate gauge, with the transformations generated by $\alpha_{\pm}$ introduced above) at the corresponding axis $\cos\theta=\pm1$, where they should be regular. These read
\begin{equation}
    Y^{\pm}_{q,\ell,m}=\Theta_{q,\ell,m}(\theta)e^{i\left(m\pm q\right)\phi}\, ,
\end{equation}
so, unless $m\pm q=0$, we have that $\Theta_{q,\ell,m}(\theta)$ must vanish at $\cos\theta=\pm1$ in order to make $Y^{\pm}_{q,\ell,m}$ regular there. On the other hand, if $m\pm q=0$ then it is enough that $\Theta_{q,\ell,m}(\theta)$ is finite at $\cos\theta=\pm1$  to make $Y^{\pm}_{q,\ell,m}$ regular. In fact, it is easy to show using \eqref{Yexp} and the properties of Jacobi polynomials that monopole harmonics with $m=-q$ and $m=+q$ do not vanish at $\cos\theta=+1$ and $\cos\theta=-1$, respectively, so we dub them the \textit{north and south monopole modes}. In particular, for the lowest-order north and south monopole modes $\ell=\lvert q\lvert$ and $m=\mp q$ we find
\begin{equation}
Y^{\pm}_{q,\lvert q \rvert,\mp q}\sim(1\pm \cos \theta)^{\lvert q \rvert}\, ,
\end{equation}
which are localised at the north and south poles. As explained in the main text, the most superradiantly unstable modes exhibit this structure, with some smooth deformation due to the black hole rotation.

\section{The Dyonic Kerr--Newman Black Hole} \label{A2}

The line element is 
\begin{equation}
\begin{aligned}\label{KN}
&ds^{2}=-\frac{\Delta-{a^2}\sin^{2}\theta}{\Sigma}dt^{2}-2a\sin^{2}\theta\left(\frac{r^2+a^2-\Delta}{\Sigma}\right)dtd\phi\\
&+\left(\frac{(r^2+a^2)^2 -\Delta a^2\sin^{2}\theta}{\Sigma}\right)\sin^{2}\theta d\phi^{2}+\frac{\Sigma}{\Delta}dr^{2}+\Sigma d\theta^{2}\, , 
\end{aligned}
\end{equation}
with $\Delta=r^{2}-2Mr+a^{2}+Q^{2}+P^{2}$ and $\Sigma=r^{2}+a^{2}\cos^{2}\theta$, while the Maxwell potential in the \textit{intermediate gauge} is
\begin{align}\label{MaxVec}
    A=&-\frac{Q r}{\Sigma}\left(dt-a \sin^{2}\theta d\phi\right)+\frac{P\cos\theta}{\Sigma} \left(a dt-(r^2+a^2)d\phi\right)\, .
\end{align}
Here, $a=J/M$ where $M$ and $J$ are the ADM mass and angular momentum, while $Q$ and $P$ are the black hole's electric and magnetic charge, defined as
\begin{equation}
Q=\frac{1}{4\pi}\int_{S^{2}}\star F\, , \ \ \ P=\frac{1}{4\pi}\int_{S^{2}} F\, ,
\end{equation}
where $S^{2}$ is any surface of constant $r$. As long as $M\geq\sqrt{a^{2}+P^{2} +Q^{2}}$, the solution exhibits Killing horizons at
\begin{equation}
   r_{\pm}=M\pm\sqrt{M^{2}-a^{2}-P^{2} -Q^{2}}\, ,
\end{equation}
where $\Delta(r_{\pm})=0$. The outermost horizon $r_{+}$ is also an event horizon and is generated by $k\equiv\partial_{t}+\Omega \partial_{\phi}$, where the angular velocity $\Omega$ and the surface gravity $\kappa$ are given by
\begin{equation}
    \Omega=\frac{a}{r_{+}^{2}+a^{2}}\,, \ \ \ \kappa=\frac{r_{+}-r_{-}}{2\left(r_{+}^{2}+a^{2}\right)}\,.
\end{equation}
We call the gauge of \eqref{MaxVec} intermediate because it is not regular at neither $\cos\theta=\pm1$ (if $P\ne0$, just like the intermediate gauge of Dirac's monopole) nor at $r=r_{+}$, as we shall discuss next.\footnote{However, we recall that all that matters is that the fieldstrength $F=dA$ is regular everywhere, as is the case here.} However, it has the advantage that its components evaluated at the horizon give the black hole electric potential $\Phi$ appearing in the first law of black hole mechanics (see e.g. \cite{Ortin:2022uxa} and the appendix of \cite{Dyson:2023ujk})
\begin{equation}\label{pot}
    \Phi=-A_{t}\left(r_{+}\right)-\Omega A_{\phi}\left(r_{+}\right)=\frac{r_{+} Q}{r^{2}_{+}+a^{2}}\, .
\end{equation}
To see that the intermediate gauge \eqref{MaxVec} is not regular at the future event horizon one can move to advanced EF coordinates,
\begin{equation}
dv=dt+\frac{r^{2}+a^{2}}{\Delta}dr\, , \ \ \     d\chi=d\phi+\frac{a}{\Delta}dr\, , \label{EF}
\end{equation}
and see immediately that $A$ diverge as $A\sim dr/(r-r_{+})$ close to the horizon $r\to r_{+}$. Of course, all singularities of $A$ can be removed by a suitable gauge transformation. In particular, we shall introduce three more gauges, $A_{\pm}$ and $A_{H}$, where the vector potential is well defined at $\cos\theta=\pm1$ and $r=r_{+}$, respectively, and given by 
\begin{align}
    A_{\pm}=A\pm P d\phi\, ,\ \  A_{H}=A-\frac{Q r}{\Delta(r)} dr\, .
\end{align}
These correspond to applying a gauge transformation on \eqref{MaxVec} generated by the local functions
\begin{align}
    \alpha_{\pm}=\mp e P \phi \,, \ \ \alpha_{H}=e\int^{r}dr \frac{Q r}{\Delta(r)} \, . \label{alphas}
\end{align}
%
\section{Flux Formulas from Covariant Phase Space} \label{A3}

Let $(g,F)$ be an electrovacuum solution, and assume $X$ is a Killing vector that also leaves invariant the Maxwell fieldstrength, so $\pounds_{X}g=0$ and $\pounds_{X}F=0$. Then it has associated a function $\mathcal{P}_{X}$ dubbed the \textit{momentum map of $X$}, defined by the equations \cite{Elgood:2020svt,Ortin:2022uxa,Dyson:2023ujk}
\begin{equation}
    d\mathcal{P}_{X}=-\iota_{X}F\, , \ \ \ \ \int_{S_{\infty}^{2}}\mathcal{P}_{X}d\Omega=0\, ,
\end{equation}
where $S_{\infty}^{2}$ denotes a 2-sphere at infinity. In the case of the dyonic KN solution, the momentum maps associated to $\partial_{t}$ and $\partial_{\phi}$ are
\begin{equation}\label{MM}
\mathcal{P}_{t}=-\frac{Q r - a P \cos \theta }{\Sigma }\, ,\ \ \ \  \mathcal{P}_{\phi}=\frac{a Q r    \sin ^2\theta-P \left(a^2+r^{2}\right)\cos \theta }{\Sigma }\, .
\end{equation}
In the special case of the Killing vector that generates the horizon, $k=\partial_{t}+\Omega\partial_{\phi}$, the associated momentum map $\mathcal{P}_{k}=\mathcal{P}_{t}+\Omega \mathcal{P}_{\phi}$ is constant at the horizon (i.e.~it satisfies a zeroth law of black hole mechanics) and its value is related to the black hole's electric potential as $\mathcal{P}_{k}(r=r_{+})=-\Phi$ \cite{Elgood:2020svt,Ortin:2022uxa,Dyson:2023ujk}. 

\begin{figure}[t!]
  \includegraphics[width=8.6cm]{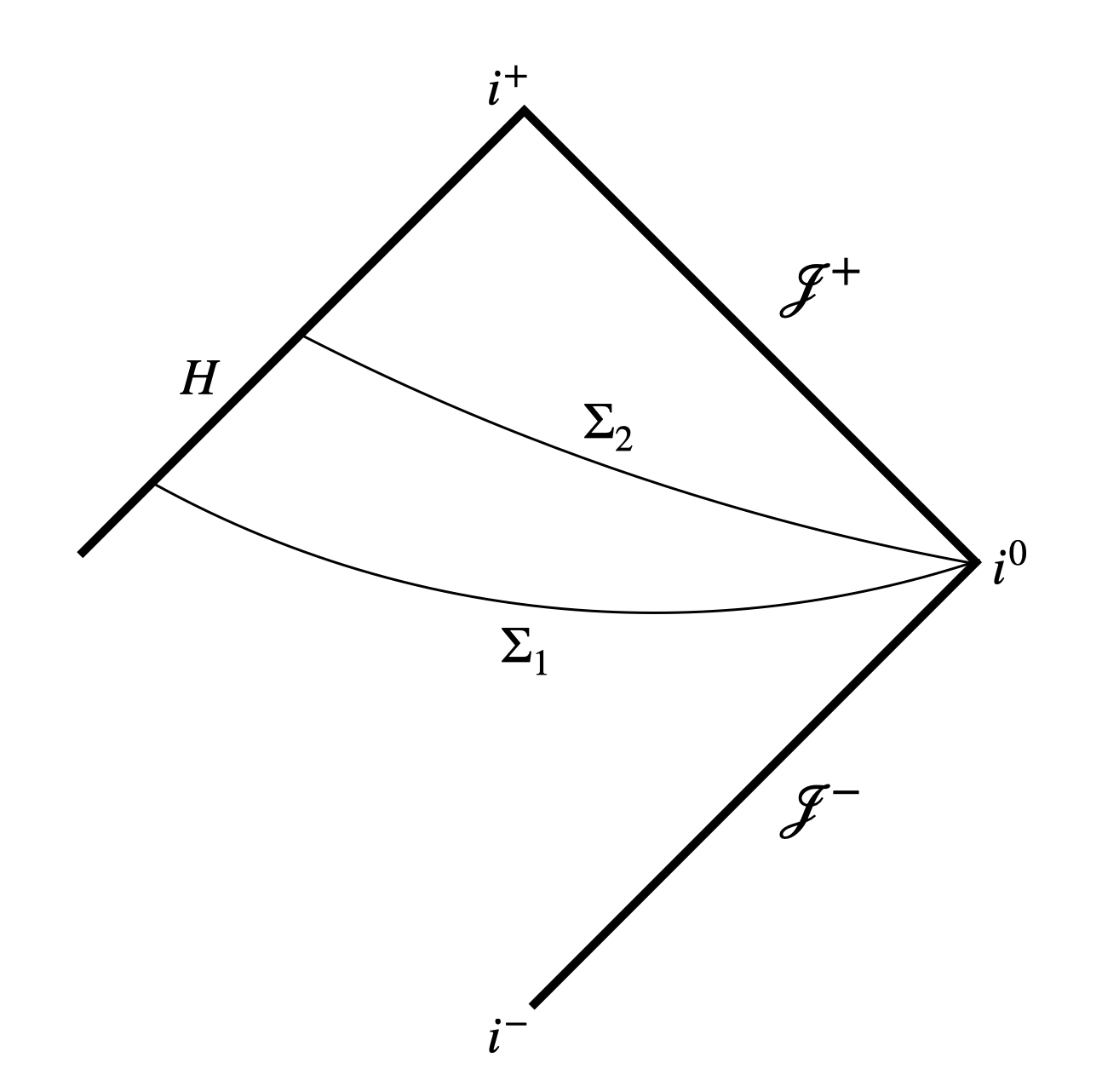}
\caption{\footnotesize{Penrose--Carter diagram illustrating our flux computation.}}\label{Sigmas}
\end{figure}

Consider now a fluctuation $\delta g, \delta F$ satisfying the Einstein--Maxwell equations linearised around some electrovacuum background solution,
\begin{equation}\label{pertfixed}
\begin{aligned}
\delta(G_{ab}-T^{F}_{ab})&=8\pi T_{ab}\, ,\\
d\delta \star F&=-4\pi e \star j\, ,
\end{aligned}
\end{equation}
where $j_{a}$ and $T_{ab}$ are generic sources and $T^{F}_{ab}$ is the Maxwell energy-momentum tensor. Then, given a Killing vector $X$ of the background, it can be shown \cite{Dyson:2023ujk} that\footnote{The notation $\boldsymbol{\epsilon}_{a}T^{a}$ means contraction with the first index of the volume form, while for the Hodge dual we use $\star T_{abc}=\boldsymbol{\epsilon}_{abcd}T^{d}$.}
\begin{equation}\label{DIFMAST}
d \bold{k}_{X}=-\mathcal{P}_{X}e\star j+X_{a}T^{ab}\boldsymbol{\epsilon}_{b}\, ,
\end{equation}
where $\bold{k}_{X}$ is the \textit{surface charge of X} \cite{Barnich:2001jy,Ortin:2022uxa,Barnich:2007bf}, which explicitly reads
\begin{equation}
    \bold{k}_{X}=\delta\bold{Q}_{X}^{GR}+\iota_{X}\boldsymbol{\Theta}^{GR}+\frac{1}{4\pi}\mathcal{P}_{X}\delta \star F\, ,
\end{equation}
where 
\begin{equation}
    \bold{Q}^{GR}_{X}=\frac{1}{16\pi}\star dX\, , \ \ \ \ \ \ \boldsymbol{\Theta}^{GR}=-\frac{1}{16\pi}\boldsymbol{\epsilon}^{a}\left(\delta^{c}_{a}\delta^{d}_{b}-g_{ab}g^{cd}\right)\delta \tensor{\Gamma}{^b _c_d}\, .
\end{equation}
The ADM energy and angular momentum are nothing but the integrals of the surface charges associated to the timelike and rotational Killing vectors $t^{a}\equiv\left(\partial_{t}\right)^{a}$ and $\phi^{a}\equiv\left(\partial_{\phi}\right)^{a}$ \cite{Iyer:1994ys,Barnich:2001jy}. Indeed, if $\Sigma$ is an asymptotically flat end, then
\begin{align}
    \delta M_{ADM}=-\int_{S_{\infty}^{2}}\bold{k}_{t}\,,  \ \ \ \ \delta J_{ADM}=\int_{S_{\infty}^{2}}\bold{k}_{\phi} \, ,
\end{align}
where $S_{\infty}^{2}$ denotes a 2-sphere at infinity. From this, together with equation \eqref{DIFMAST} and Stokes' theorem it follows immediately that the energy and angular momentum contained in an hypersurface $\Sigma$ (at first order in the fluctuation) are
\begin{align}
    E(\Sigma)= \int_{\Sigma}\left(\mathcal{P}_{t}e\star j-t_{a}T^{ab}\boldsymbol{\epsilon}_{b}\right)\, , \ \ \ \ J(\Sigma)=  \int_{\Sigma}\left(-\mathcal{P}_{\phi}e\star j+\phi_{a}T^{ab}\boldsymbol{\epsilon}_{b}\right)\,. 
\end{align}
In parallel, the electric charge in $\Sigma$ is
\begin{equation}
    Q(\Sigma)=-e\int_{\Sigma}\star j\,.
\end{equation}
Assume now that the background spacetime has an event horizon which is a Killing horizon of the vector field $k=\partial_{t}+\Omega\partial_{\phi}$, where $\Omega$ is a constant. We want to obtain the flux of energy, angular momentum and charge through a portion $H$ of the horizon, defined by its intersection with two spacelike hypersurfaces $\Sigma_{1}$ and $\Sigma_{2}$, with $\Sigma_{2}$ lying into the future of $\Sigma_{1}$ (see Figure \ref{Sigmas}). By Stokes' theorem we have
\begin{equation}
    E(\Sigma_{1})=E(\Sigma_{2})+E(H)\, ,
\end{equation}
and therefore the amount of energy flowing through the horizon is
\begin{equation}\label{DelEAp}
\Delta E=-\left(E(\Sigma_{2})-E(\Sigma_{1})\right)=E(H)=\int_{H}\left(\mathcal{P}_{t} e \ j^{a}k_{a}+t^{a}T_{ab}k^{b}\right)\tilde{\boldsymbol{\epsilon}}
\end{equation}
where in the last step we used that the volume form on $H$ relative to the horizon generator $k^{a}$, denoted $\tilde{\boldsymbol{\epsilon}}$, is related to the total volume form $\boldsymbol{\epsilon}$ by (see e.g.~\cite{Gao:2001ut})
\begin{equation}
\boldsymbol{\epsilon}_{abcd}=-4k_{[a}   \tilde{\boldsymbol{\epsilon}}_{bcd]}\, .
\end{equation}
The same argument can be applied to compute the fluxes of angular momentum and charge, which give
\begin{align}\label{DelJAp}
    \Delta J=&-\int_{H}\left(\mathcal{P}_{\phi} e \ j^{a}k_{a}+\phi^{a}T_{ab}k^{b}\right)\tilde{\boldsymbol{\epsilon}}\, ,\\ \label{DelQAp}
    \Delta Q=&-e\int_{H}j^{a}k_{a} \ \tilde{\boldsymbol{\epsilon}}\, .
\end{align}
Equations \eqref{DelEAp}, \eqref{DelJAp} and \eqref{DelQAp} give the fluxes of energy, angular momentum and charge induced by a general source of the gravitational and electromagnetic fields. Given a model for the source these can be evaluated straightforwardly. In this work the source is a charged, complex scalar field, and we have
\begin{align}
     j_{a}=\frac{i}{2}\left(\bar{\varphi}D_{a}\varphi-\varphi D_{a}\bar{\varphi}\right)\, , \ \ \ T_{ab}=\frac{1}{2} \left[D_{a}\bar{\varphi}D_{b}\varphi +D_{a}\varphi D_{b}\bar{\varphi}-\left(D_{c}\bar{\varphi}D^{c}\varphi+V(\bar{\varphi}\varphi)\right)g_{ab}\right]\, ,
\end{align}
where $V(\bar{\varphi}\varphi)=\mu^{2}\bar{\varphi}\varphi$. Furthermore, we want to evaluate the fluxes induced by a mode of the form
\begin{equation}\label{modeAp}
\varphi=e^{-i\omega v}e^{i m\chi}\psi(r,\theta)\, ,
\end{equation}
with $\omega=\omega_{R}+i \omega_{I} $, and taking the dyonic KN solution as a background (notice that one should not worry about gauge ill-definiteness of $\varphi$ at $H$ since the quantities we are computing are gauge-invariant). A straightforward substitution of \eqref{modeAp} into \eqref{DelEAp}, \eqref{DelJAp} and \eqref{DelQAp} yields the flux formulas in the main text. 
%

\section{Leaver's Method} \label{A4}

Writing $\varphi=R(r)S(\theta)e^{-i\omega t+ i m \phi}$ (again in the intermediate gauge), the equation of motion for the scalar field reduces to
    \begin{align}
    &\frac{1}{\sin\theta}\frac{d}{d\theta}\left(\sin\theta \frac{d}{d\theta}S(\theta)\right)-\left(\frac{(m+q\cos\theta-a\omega\sin^{2}\theta)^{2}}{\sin^{2}\theta}+a^{2}\mu^{2}\cos^{2}\theta-\Lambda_{q\ell m}+q^{2}+2a m\omega-a^{2}\omega^{2}\right)S(\theta)=0\, , \\
    &\frac{d}{dr}\left(\Delta(r)\frac{d}{dr}R(r)\right)+\left(\frac{\left(\left(r^{2}+a^{2}\right)\omega-m a +eQr\right)^{2}}{\Delta(r)}-\mu^{2}r^{2}-\Lambda_{q\ell m}+q^{2}+2am\omega-a^{2}\omega^{2}\right)R(r)=0\, ,
\end{align}
where $\Lambda_{q\ell m}$ is a separation constant. We notice that for $a=0$ the angular equation reduces precisely to that of monopole spherical harmonics in Appendix \ref{A1}, and in that case $\Lambda_{q \ell m}=\ell(\ell+1)$. Likewise, for $q=0$ and $Q=0$ the separation constant reduces to $\Lambda_{\ell m}$ as defined in \cite{Dolan:2007mj}. Introducing the variable $x=\cos\theta$, the behaviour of $S(x)$ close to the poles $x=\pm1$ is
\begin{equation}\label{behS}
    S(x)\sim\begin{cases}(1-x)^{\pm\frac{m+q}{2}}\, , & x\to1 \\ (1+x)^{\pm\frac{m-q}{2}}\, , & x\to-1\end{cases}\, .
\end{equation}
Regarding the radial function $R(r)$, the behaviour at the horizon $r_{+}$ and infinity $r=\infty$ is
\begin{equation}\label{behR}
    R(r)\sim\begin{cases}(r-r_{+})^{\pm i\frac{\tilde{\omega}}{2\kappa}}\, , & r\to r_{+} \\  e^{\pm i \sqrt{\omega^{2}-\mu^{2}}r}\, , & r\to\infty\end{cases}
\end{equation}
where $\tilde{\omega}=\omega-m\Omega+e\Phi$ (we notice it differs from $\omega_{*}=\omega_{R}-m\Omega+e\Phi$ in the imaginary part of $\omega$, that is, $\omega_{I}$). We look for solutions that are regular as $U(1)$ sections on and outside the horizon. The fields $\varphi_{\pm}$ and $\varphi_{H}$ (obtained after applying the gauge transformations generated by \eqref{alphas}) read
\begin{align}
    \varphi_{\pm}&=e^{-i\omega t}e^{i(m\pm q)\phi}R(r)S(x)\, , \\
    \varphi_{H}&\sim e^{-i\omega v}e^{i m\chi}(r-r_{+})^{i\frac{\tilde{\omega}}{2\kappa}}R(r)S(x)\, ,
\end{align}
where the last expression, written in advanced EF coordinates, holds only close to the horizon $r\to r_{+}$. Regularity of $\varphi$ as a section implies that $\varphi_{\pm}$ and $\varphi_{H}$ must be regular at $\cos\theta=\pm1$ and $r=r_{+}$, respectively, and this fixes which behaviours in \eqref{behS}-\eqref{behR} must be chosen for $R(r)$ and $S(x)$. The boundary conditions at infinity depend on the class of modes we are looking for. For quasinormal modes (QNMs) and bound states (BSs), at infinity one only allows purely outgoing waves, and exponential suppression, respectively. This makes the problem over-determined, and solutions will exist only for some values of $\omega$ and $\Lambda_{q\ell m}$. 

Following Leaver's method \cite{Leaver:1985ax,Cardoso:2005vk,Dolan:2007mj}, we propose an ansatz that has automatically implemented our boundary conditions at the horizon and at $x=-1$,
 \begin{align}\label{s}
    S(x)&=e^{a\sqrt{\omega^{2}-\mu^{2}}x}(1+x)^{\lvert \frac{m-q}{2}\rvert}(1-x)^{\lvert \frac{m+q}{2}\rvert}\sum_{k=0}^{\infty}a_{k}(1+x)^{k}\, ,\\ \label{r}
    R(r)&=e^{i\sqrt{\omega^{2}-\mu^{2}}r}(r-r_{-})^{i\frac{\tilde{\omega}}{2\kappa}-i\frac{M(\mu^{2}-2 \omega^{2})-e Q \omega}{\sqrt{\omega^{2}-\mu^{2}}}-1}\left(r-r_{+}\right)^{-i\frac{\tilde{\omega}}{2\kappa}}\sum_{k=0}^{\infty}d_{k}\left(\frac{r-r_{+}}{r-r_{-}}\right)^{k}\, ,
\end{align}   
where $a_{k}$ and $d_{k}$ are constants subject to the following three-term recurrence relation,
\begin{equation}
  \begin{aligned}\label{reca}
       & \alpha^{\theta}_{0}a_{1}+\beta^{\theta}_{0}a_{0}=0\,\\
        &\alpha^{\theta}_{k}a_{k+1}+\beta^{\theta}_{k}a_{k}+\gamma^{\theta}_{k}a_{k-1}=0\, , \ \ \ k>0
\end{aligned}  
\end{equation}
and
\begin{equation}
    \begin{aligned}\label{recd}
       & \alpha^{r}_{0}d_{1}+\beta^{r}_{0}d_{0}=0\,\\
        &\alpha^{r}_{k}d_{k+1}+\beta^{r}_{k}d_{k}+\gamma^{r}_{k}d_{k-1}=0\, , \ \ \ k>0
\end{aligned}
\end{equation}
where $\alpha^{r,\theta}_{k}$, $\beta^{r,\theta}_{k}$ and $\gamma^{r,\theta}_{k}$ can be found below in Appendix \ref{A7}. The QNM/BS boundary conditions at infinity $r=\infty$ as well as regularity at $x=1$ will also be satisfied if the previous series converge there. This can be achieved by supplementing \eqref{reca}-\eqref{recd} with an additional ``asymptotic'' boundary condition\footnote{This additional condition, which in essence is requiring the convergence of an infinite continued fraction, implies by Pincherle's theorem that our solution to the recurrence relation is \textit{minimal}, which is the right choice if we want the sums $\sum_{k}d_{k}$ and $\sum_{k}a_{k}2^{k}$ to converge (see \cite{Leaver:1985ax} and references therein).}
\begin{equation}
\frac{a_{1}}{a_{0}}=\frac{-\gamma^{\theta}_{1}}{\beta^{\theta}_{1}-}\frac{\alpha^{\theta}_{1}\gamma^{\theta}_{2}}{\beta^{\theta}_{2}-}\frac{\alpha^{\theta}_{2}\gamma^{\theta}_{3}}{\beta^{\theta}_{3}-}...\, , 
\end{equation}
and
\begin{equation}
\frac{d_{1}}{d_{0}}=\frac{-\gamma^{r}_{1}}{\beta^{r}_{1}-}\frac{\alpha^{r}_{1}\gamma^{r}_{2}}{\beta^{r}_{2}-}\frac{\alpha^{r}_{2}\gamma^{r}_{3}}{\beta^{r}_{3}-}...\, .
\end{equation}
This additional condition makes the problem over-determined, and will only have solutions for specific values of $\Lambda_{q\ell m}$ and $\omega$. Then, QNMs and BSs are defined as solutions with:
\begin{align}
    \text{QNM:}&  \ \ \ \ (\sqrt{\omega^{2}-\mu^{2}})_{I}<0\, ,\\
    \text{BS:}&\ \ \ \ (\sqrt{\omega^{2}-\mu^{2}})_{I}>0\, .
\end{align}
where the subscript $I$ stands for the imaginary part.
\section{Superradiant Amplification and Energy Extraction} \label{A5}

As a consistency check of our interpretation of the instability, here we briefly discuss the superradiant amplification factors for massless fields, together with the behaviour of the fluxes of energy, angular momentum and charge of massive bound states. To compute the amplification factors it is convenient to define a new radial function $\Psi(r)=\frac{R(r)}{\sqrt{r^2+a^2}}$ that satisfies a Schr\"odinger-like equation
\begin{equation}\label{eq:schro_like}
\left(\frac{d^2}{d r_{*}^2}+V(r)\right)\Psi\left(r_{*}\right)=0\,,
\end{equation}
where $r_*$ is a tortoise coordinate defined through $dr_*/dr=(r^2+a^2)/\Delta$. The effective potential $V(r)$ is given by
\begin{equation}
    V(r)=\frac{H(r)^2+\left(q^2-\Lambda_{q
\ell m}+2 m a \omega-\mu^2 r^2-a^2\omega^2\right) \Delta}{\left(r^2+a^2\right)^2}-\frac{\Delta[\Delta+2 r(r-M)]}{\left(r^2+a^2\right)^3}+\frac{3 r^2 \Delta^2}{\left(r^2+a^2\right)^4}\,,
\end{equation}
where we defined $H(r)\equiv\left(r^2+a^2\right) \omega-a m-e Q r$. The solution $\Psi(r)$ admits the following asymptotic solutions at the horizon $r_*\to -\infty$ and infinity $r_*\to +\infty$
\begin{equation}
    \Psi(r_*)\sim\begin{cases}\mathcal{T} e^{-i\omega_* r_*}\, , & r_*\to -\infty \\  \mathcal{I} e^{-ik_{\infty}r_*}r^{-i\nu}+\mathcal{R} e^{+ik_{\infty}r_*}r^{i\nu}\, , & r_*\to+\infty\end{cases}
\end{equation}
where $\omega_*=\omega-m\Omega+e\Phi$, $k_{\infty}=\sqrt{\omega^{2}-\mu^{2}}$ and $\nu=i(M\mu^2+e Q\omega)/k_{\infty}$. Assuming $\omega>\mu$ for $\omega\in \mathbb{R}$, these solutions describe the problem of a wave with amplitude $\mathcal{I}$ scattering off the potential barrier, part of it being reflected back with reflection coefficient $\mathcal{R}$, and another part being partially transmitted through the potential barrier and into the event horizon, with transmission coefficient $\mathcal{T}$. As follows immediately from the flux formulas in the main text, whenever $\omega_*<0$ (assuming $\omega\geq 0$) energy is being extracted from the hole and one should expect $|\mathcal{R}|^2>|\mathcal{I}|^2$~\cite{Brito:2015oca}, that is, the wave is \emph{superradiantly amplified} for sufficiently small frequencies $\omega$.

In what follows we define the amplification factor as
\begin{equation}
Z=\frac{|\mathcal{R}|^2}{|\mathcal{I}|^2}-1
\end{equation}
such that $Z>0$ whenever superradiant scattering occurs. In order to compute the amplification factor we modified the \texttt{Mathematica} notebooks publicly available at~\cite{grit_website}. For a chosen value of $\omega$, we first compute the eigenvalue $\Lambda_{q\ell m}$ using Leaver's method and then numerically integrate Eq.~\eqref{eq:schro_like} for each value of $\omega$, using the asymptotic solution at the event horizon as an initial condition. The integration is done up to an arbitrarily large value of $r=r_{\infty}$ where, to extract the coefficients $\mathcal{I}$ and $\mathcal{R}$, we match the solution and its derivative to the asymptotic solution at infinity. Typically $r_{\infty}=10^3/|\omega|$ is enough to achieve very good convergence of the results. As usual, to reduce truncation errors we also expand the asymptotic solutions at the event horizon and at infinity in a series expansion (as described in e.g. Ref.~\cite{Benone:2019all}). 

\begin{figure}[t!]
  \includegraphics[width=8cm]{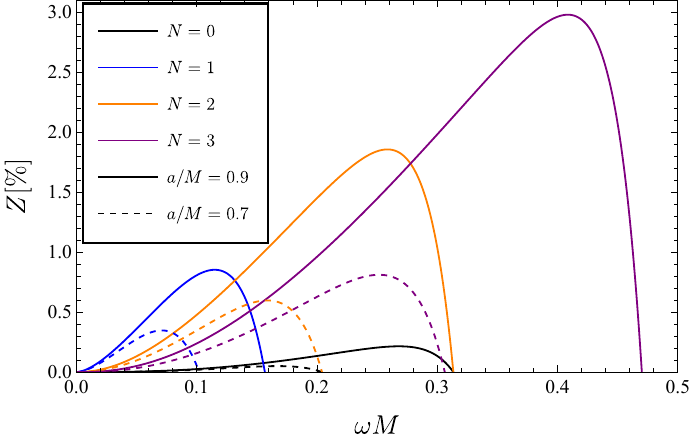}
\caption{Amplification factor as a function of the frequency $\omega$ of a massless wave ($\mu=0$) scattered off a magnetic black hole with $N=0,1,2,3$ monopoles spinning at $a/M=0.7$ and $a/M=0.9$. For $N=0$ (corresponding to the neutral Kerr black hole case) we show the mode $\ell=m=1$, while for $N=1,2,3$ we show the north monopole modes $\ell=m=-q=N/2$. The black hole magnetic charge is fixed to be arbitrarily small $P\sim 0$. The amplification factors for the south monopole mode can be obtained from the north monopole ones by $\omega\to -\omega$.}\label{fig:super}
\end{figure}
\begin{figure}[t!]
  \includegraphics[width=18cm]{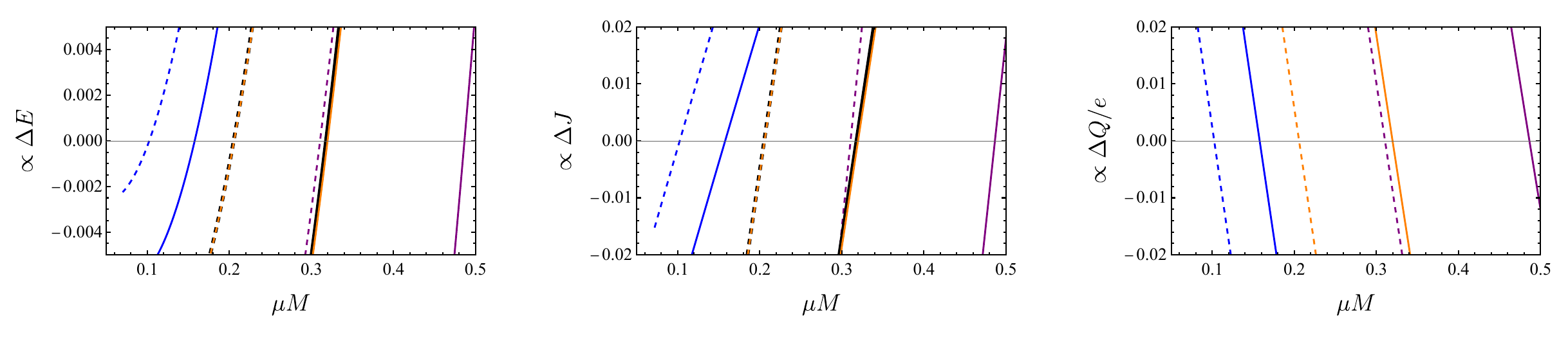}
\caption{Fluxes of energy, angular momentum and charge across the black hole horizon, for $a,P,e>0$, focusing on the north monopole mode $\ell=m=-q=N/2$. The plots show a quantity proportional to the fluxes, so they only capture their sign, and not the magnitude. The energy and angular momentum fluxes remain invariant for the south mode, but the charge flux changes sign. The color and curve styles follow the legend of Fig.~\ref{fig:super}.}\label{fluxes}
\end{figure}

Our results are shown in Fig.~\ref{fig:super} where we consider a massless wave ($\mu=0$) scattering off a magnetic black hole with $N=1,2,3$ monopoles. The black hole is assumed to have zero electric charge $Q=0$ and we set the magnetic charge to an arbitrarily small value $P/M\sim 0$, such that the results are independent of $P$ (and so they are comparable to the regime of parameters considered in the text, where $P/M\sim10^{-19}$). We compare the amplification of a mode $\ell=m=1$ scattering off a neutral Kerr black hole (corresponding to the $N=0$ case), against the amplification of north monopole modes with $\ell=m=-q=N/2$. We do not show the amplification factors for the south monopole mode since those can be obtained from the north monopole ones by switching $\omega\to -\omega$.
As expected, we obtain that the amplification factors become positive precisely when $\omega_{*}$ turns negative. Just as it happens for the instability timescale, we obtain a significant increase in the amplification of superradiant modes with respect to the neutral Kerr case even for a low number of magnetic monopoles and a negligible magnetic charge. 

It is also instructive to evaluate the fluxes of energy, angular momentum and charge for the massive bound states (notice these are different form the massless modes just considered in computing the amplification factors). Focusing only on their sign, i.e. just evaluating the prefactors in the flux formulas, we obtain the results shown in Fig.~\ref{fluxes}. When compared with the instability plot (Fig.\ref{instabs} in the main text), we observe that the mode starts extracting energy and angular momentum precisely at the point that it becomes unstable. For $a,P,e>0$ we obtain that the north (south) mode starts inducing positive (negative) charge into the hole when it becomes unstable, thus growing a negative (positive) cloud in the exterior. All these facts together confirm our picture of the superradiant instability of magnetic black holes presented in the concluding discussion of the manuscript.
\section{QNMs and Higher Modes} \label{A6}

\begin{figure}[t!]
  \includegraphics[width=8cm]{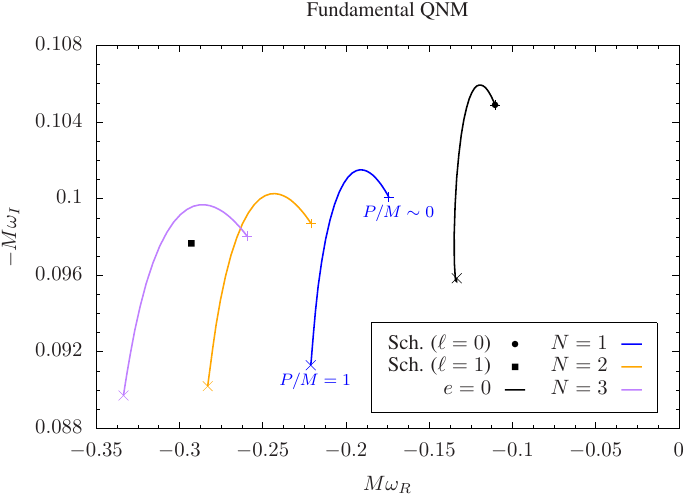}
  \includegraphics[width=8cm]{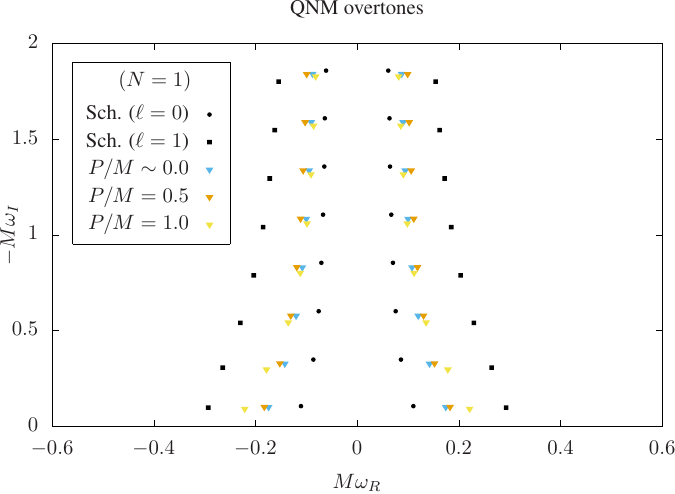}
\caption{ QNM for massless scalar field ($\mu=0$) propagating on static ($a=0$) magnetic black hole. The circle and square points correspond, respectively, to the $\ell=0$ and $\ell=1$ QNMs modes in Schwarzschild. {\em Left Panel:} Fundamental QNM for magnetic black holes with $N=1,2,3$ monopoles, varying the magnetic charge within $P/M\in(0,1]$. The limit $P/M\to0$ and the extremal case $P/M=1$ are marked by `$+$' and a `$\times$' signs, respectively. The black curve corresponds to a massless scalar field with electric charge $e=0$. {\em Right Panel:} QNM overtones for a black hole with $N=1$ monopole, and magnetic charges $P/M\sim0$, $P/M=0.5$ and $P/M=1$. These QNMs have been computed with hyperboloidal methods\cite{Jaramillo:2020tuu,PanossoMacedo:2023qzp}, matching accurately the results obtained using the Leaver's method employed for the rest of computations in the paper.}
\label{fig:QNM_massless_a0}
\end{figure}

\begin{figure}[t!]
  \includegraphics[width=8cm]{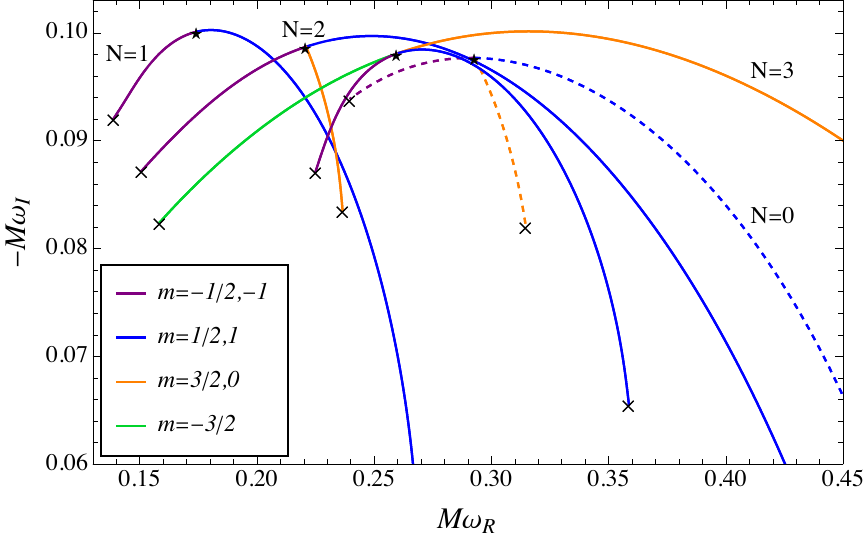}
  \includegraphics[width=8cm]{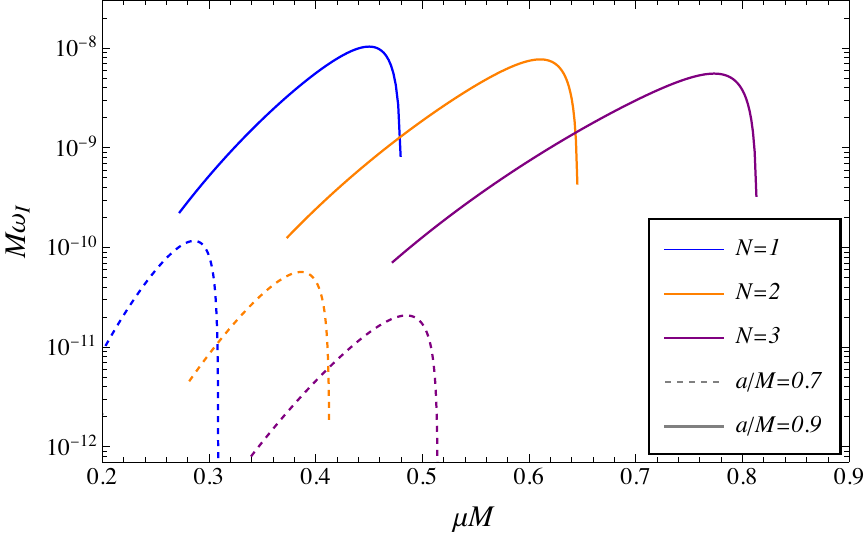}
\caption{{\em Left Panel:} QNMs for massless scalar field on a magnetic Kerr-Newman black hole, containing $N=0, 1, 2, 3$ magnetic monopoles. Black stars correspond the fundamental frequencies of the modes $\ell=N/2$ at zero black hole spin, $a=0$, which do not depend on $m$. As $a$ increases from 0 to 0.95, from the black stars emanate the frequencies for the modes $m=-N/2,-N/2+1,...,N/2$ (colored as indicated in the legend) where the end point is indicated with a cross. To ease comparison, the neutral case $N=0$ is represented with dashed lines. {\em Right Panel:} Superradiant instability growth rate of the bound states for a charged field on a magnetic Kerr-Newman black hole background. We assumed for the field the charge-to-mass ratio of charged pions, and considered black holes with $N=1,2,3$ magnetic monopoles, spinning at $a=0.7$ and $a=0.9$. The results are relative to the higher modes with $\ell=|q|+1$ and $m=\ell$.}
\label{fig:QNMs_Rotating_Leaver-Instabilities_higher_modes}
\end{figure}

For completeness, here we report some results for the QNMs of a charged, massless scalar field evolving on a magnetic Kerr-Newman black hole (see Fig.~\ref{fig:QNM_massless_a0} and left panel of Fig.~\ref{fig:QNMs_Rotating_Leaver-Instabilities_higher_modes}). As mentioned in the main text, all QNMs are stable and their structure resembles that of a scalar field on a Kerr black hole. However, we again observe differences of order one even if the black hole contains a small number of monopoles. In particular, for $N\ne0$, the QNMs do not approach those of a neutral scalar field in Kerr in the limit $P/M\to0$ (see Fig.~\ref{fig:QNM_massless_a0}). Finally, we also considered the superradiant instability of some higher modes $\ell=\lvert q\rvert +1$ (see right panel of Fig.~\ref{fig:QNMs_Rotating_Leaver-Instabilities_higher_modes}), and found that these are less unstable than the modes $\ell=\lvert q \rvert$. We have checked that the same is true for other higher modes.
%

\section{Coefficients in Recurrence Relations}\label{A7}

Setting $M=1$ and defining $\mathcal{D}=\sqrt{1-a^{2}-P^{2}-Q^{2}}$, we have, for the radial coefficients,
\begin{align}
    \alpha^{r}_{k}&=k^{2}+(1+A_{0})k+A_{0}\, , \\
    \beta^{r}_{k}&=-2 k^{2}+(2+A_{1})k+B_{0}\, , \\
    \gamma^{r}_{k}&=k^{2}+(A_{2}-3)k+B_{1}-A_{2}+2
\end{align}
where
\begin{align}
    A_{0}&=-\frac{i \left(a^2 \omega -a m+\mathcal{D}^2 \omega +\mathcal{D} (2 \omega +i)+(\mathcal{D}+1) e Q+\omega \right)}{\mathcal{D}}\, , \\
    A_{1}&=-2(1+A_{0})-2i\frac{  (2 \mathcal{D}+1) \mu ^2-\omega  (2 (\mathcal{D}+1) \omega +e Q)}{\sqrt{\omega ^2-\mu ^2}}\, , \\
    A_{2}&=2+A_{0}-2i\frac{ \omega  (e Q+2 \omega )-\mu ^2}{\sqrt{\omega ^2-\mu ^2}}\, , \\ \notag
    B_{0}&=\frac{(2 \omega +i) \left(a^2 \omega -a m+\omega \right)}{\mathcal{D}}+\frac{e Q}{\mathcal{D}} \left(a^2 \omega -a m+3 \mathcal{D}^2 \omega +\mathcal{D} (6 \omega +i)+3 \omega +i\right)\\ \notag
    &+\left(a^2 \omega ^2-\mu ^2+q^2+6 \omega ^2+2 i \omega -\Lambda_{q\ell m}-1\right)-\mathcal{D}^2 \left(\mu ^2-2 \omega ^2\right)+\mathcal{D} \left(-2 \mu ^2+\omega  (6 \omega +i)\right)+\frac{(\mathcal{D}+1) e^2 Q^2}{\mathcal{D}}\\
    &-i A_{0}\frac{ (2 \mathcal{D}+1) \mu ^2-\omega  (2 (\mathcal{D}+1) \omega +e Q)}{\sqrt{\omega ^2-\mu ^2}}\, , \\
    B_{1}&=\frac{B_{2}}{\mathcal{D} \left(\mu ^2-\omega ^2\right)}-i(A_{0}+1)\frac{\omega  (e Q+2 \omega )-\mu ^2}{\sqrt{\omega ^2-\mu ^2}}\, , \\ \notag
    B_{2}&=e Q \left(\omega ^2 \left(a^2 \omega -a \mathit{m}+\mathcal{D}^2 \omega +4 \mathcal{D} \omega +i \mathcal{D}+3 \omega +i\right)-\mu ^2 \left(a^2 \omega -a \mathit{m}+\mathcal{D}^2 \omega +2 \mathcal{D} \omega +i \mathcal{D}+3 \omega +i\right)\right)\\ \notag
    &+(2 \omega +i) \left(\omega ^2-\mu ^2\right) \left(a^2 \omega -a \mathit{m}+\omega \right)+\mathcal{D}^2 \omega  (2 \omega +i) \left(\omega ^2-\mu ^2\right)\\
    &+\mathcal{D} \left(\mu ^4+\mu ^2 \left(-4 \omega ^2-2 i \omega +1\right)+\omega ^2 \left(4 \omega ^2+2 i \omega -1\right)\right)+e^2 Q^2 \left((\mathcal{D}+1) \omega ^2-\mu ^2\right)\, .
\end{align}
For the angular coefficients we find
\begin{align}
    \alpha^{\theta}_{k}&=2 k^{2}+(2+G_{0})k+G_{0}\, , \\
    \beta^{\theta}_{k}&=-k^{2}+(1+G_{1})k+H_{0}\, , \\
    \gamma^{\theta}_{k}&=G_{2}k+H_{1}-G_{2}\, ,
\end{align}
where
\begin{align}
    G_{0}&=2 (| m-q| +1)\, , \\
    G_{1}&=4 a \sqrt{\omega ^2-\mu ^2}-| \mathit{m}-q| -| \mathit{m}+q| -2\, , \\
    G_{2}&=-2 a \sqrt{\omega ^2-\mu ^2}\, , \\ \notag
    H_{0}&=-a^2 \mu ^2+a^2 \omega ^2+2 \sqrt{\omega ^2-\mu ^2} (a | m-q| +a)-2 a q \omega+\Lambda_{q\ell m} \\ 
    &-\frac{1}{2} \left(| m-q| +| m-q|  | m+q| +| m+q| +m^2+q^2\right) \, , \\
    H_{1}&=2 a q \omega -a \sqrt{\omega ^2-\mu ^2} (| m-q| +| m+q| +2)\, .
\end{align}

\end{document}